\documentclass[aps,amsmath,amssymb,reprint,superscriptaddress,showpacs]{revtex4-1}
\usepackage{graphicx}
\usepackage[colorlinks=true,linkcolor=blue,filecolor=blue,urlcolor=blue]{hyperref}
\usepackage{epstopdf}
\pdfoutput=1 
\usepackage{color}

\newcommand{\hex}{%
\raisebox{-0.3ex}{\includegraphics[scale=0.1]{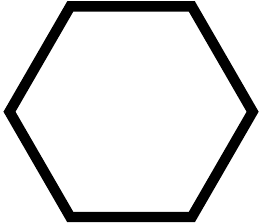}}}

\begin{document}

\title{Magnetization plateaus of an easy-axis Kagom\'e antiferromagnet with extended interactions}

\author{X. Plat}
\affiliation{Laboratoire de Physique Th\'eorique, Universit\'e de Toulouse and CNRS, UPS (IRSAMC), F-31062, Toulouse, France.}
\author{F. Alet}
\affiliation{Laboratoire de Physique Th\'eorique, Universit\'e de Toulouse and CNRS, UPS (IRSAMC), F-31062, Toulouse, France.}
\author{S. Capponi}
\affiliation{Laboratoire de Physique Th\'eorique, Universit\'e de Toulouse and CNRS, UPS (IRSAMC), F-31062, Toulouse, France.}
\author{K. Totsuka}
\affiliation{Yukawa Institute for Theoretical Physics, Kyoto University, Kitashirakawa Oiwake-Cho, Kyoto 606-8502, Japan.}

\date{\today}


\begin{abstract}
  We investigate the properties in finite magnetic field of an
  extended anisotropic XXZ spin-1/2 model on the Kagom\'e lattice, originally introduced by Balents, Fisher, and Girvin [Phys. Rev. B,
  \textbf{65}, 224412 (2002)]. The magnetization curve displays plateaus at magnetization $m=1/6$ and $1/3$ when the anisotropy is large. Using low-energy effective constrained models (quantum loop and quantum dimer models), we discuss the nature of the plateau phases, found to be crystals that break discrete rotation and/or translation symmetries.  
Large-scale quantum Monte-Carlo simulations were carried out in particular for the $m=1/6$ plateau.  
We first map out the phase diagram of the effective quantum loop model with an additional loop-loop interaction to find stripe order around the point 
relevant for the original model as well as a topological $\mathbb{Z}_{2}$ spin liquid.   
The existence of a stripe crystalline phase is further evidenced by measuring both standard structure factor 
and entanglement entropy of the original microscopic model. 
\end{abstract}

\maketitle

\section{Introduction}\label{sec:introduction}
The study of the disruption of the ordering tendency in low-dimensional antiferromagnets 
at zero temperature and the subsequent emergence of unconventional phases is an active topic of research, on both theoretical or experimental fronts. Frustrating interactions, quantum fluctuations or imposition of a magnetic field are among the main ingredients to destabilize the antiferromagnetic long-range order~\cite{FrustrationBook}. Considerable efforts have been devoted to studying and characterizing the resulting phases of matter, which may assume another type of non-magnetic ordering (such as crystalline states of singlet valence bonds), or may not exhibit any kind of local order at all ({\it quantum spin liquids}~\cite{Balents2010}). 

Two-dimensional non-bipartite lattice with antiferromagnetic interactions between spin-1/2 moments are amongst the most studied, as antiferromagnetic order is most severely challenged by quantum fluctuations there. The recent numerical findings of topological spin liquid phases in the realistic SU(2) Kagom\'e-lattice antiferromagnet~\cite{Yan2011,Depenbrock2012,Jiang2012a}, or its XXZ version~\cite{He2014b} provide examples of stabilization of such exotic phases of matter. Previously, $\mathbb{Z}_2$ topological spin liquids were found in toy models, such as constrained quantum dimer models (QDM)~\cite{Rokhsar1988} on the frustrated triangular~\cite{Moessner2001} or Kagom\'{e}~\cite{Misguich2002} lattices. These simplified models advantageously allow for a better analytical handle and understanding of the physics of spin liquids, but their connection to SU(2) microscopic Hamiltonians is not always direct.

Another route to realizing exotic phases of matter is to start from a model with highly degenerate 
manifold of ground states ({\em ice manifold}) obeying local constraints and then derive an effective Hamiltonian describing the emergent degrees of freedom in this manifold \cite{Balents2002,Senthil2002,Hermele-F-B-04}. Following this strategy, Balents, Fisher and Girvin (BFG) introduced a spin-1/2 XXZ model with extended interactions on the Kagom\'e lattice~\cite{Balents2002}. 
Using a generalized QDM derived for low energies, these authors argued that this system hosted a topological gapped $\mathbb{Z}_2$ spin liquid phase~\cite{Balents2002}, which was further confirmed numerically in subsequent works~\cite{Sheng2005,Isakov2006,Isakov2012}. One of the hallmarks of a $\mathbb{Z}_2$ liquid phase is the existence of a topological correction to entanglement entropy, which was shown to be present in the BFG model~\cite{Isakov2011}.

In considering the nature of the (gapped) phases resulting from the destruction of antiferromagnetic order, one important guiding principle is provided by a general statement about the relation between the presence/absence of an excitation gap and commensurability. For instance, it can be rigorously shown \cite{Hastings2004,Nachtergaele-S-07} that a given spin system (in the absence of magnetic field) is gapless when the value of total spin per unit cell is an half-odd-integer (this is the case for, {\em e.g.}, the spin-1/2 Heisenberg model on the Kagom\'e lattice) and the ground state is non-degenerate.  That is, if we have a featureless liquid-like ground state with an excitation gap, the state necessarily has a {\em hidden} degeneracy (probably of topological origin). A simpler version of this argument \cite{Oshikawa2000} applies to any spin-$S$ systems with U(1) symmetry and is of direct relevance to magnetization plateaux (see \cite{FrustrationBook} for a review): a unique ground state with a gap is possible only when the number of spins $q$ within a lattice unit cell and the magnetization $m$ per site satisfy $q(S-m)\in \mathbb{Z}$. Recently, a field-theoretical meaning was given to this relation~\cite{Tanaka2009} and it was predicted that when the $q(S-m)$ is a simple (non-integral) rational number, we may have gapped featureless liquid phases ({\em i.e.}, spin-liquid plateaus) as well as more conventional plateaus with magnetic superstructures. The search for microscopic models that potentially host these spin-liquid plateaus formed in high magnetic fields is quite intriguing in the light of both the field control of spin liquids \cite{Pratt-etal-11} and a recent theoretical report \cite{Nishimoto-S-H-13}.  

In this paper, we will investigate the ground-state phases of the BFG model in the presence of a magnetic field, which, as we will see, exhibits large magnetization plateaus at magnetization per site $m=1/6$ and $1/3$ as well as the one at $m=0$ (the gapped spin liquid reported in Ref.~\cite{Balents2002}). On these plateaus, the low-energy physics is also well captured by effective constrained models on a triangular lattice: a generalized QDM for the $m=0$ plateau \cite{Balents2002}, the usual QDM for $m=1/3$, and a quantum loop model (QLM) for $m=1/6$. While much is known already about the nature of the $m=1/3$ plateau thanks to numerous extensive studies~\cite{Moessner2001,Ralko2005,Syljuasen2005,Ralko2006,Vernay2006,Ralko2007,Misguich2008a,Herdman-W-11} done for the QDM on the triangular lattice, the phase diagram of the QLM, which is of direct relevance to the $m=1/6$ plateau, is not known to our best knowledge. We will investigate it in details using simple analytical considerations and quantum Monte Carlo (QMC) simulations. 

The paper is organized as follows. In Sec.~\ref{sec:model}, we present the spin 1/2 model considered in this study and provide the low-energy mapping to constrained models valid for the magnetization plateaus at $m=1/6$ (QLM) 
and $1/3$ (QDM). We also review some key results obtained previously when no magnetic field is present.  Sec.~\ref{sec:loop_model} is devoted to a complete mapping of the phase diagram of the QLM on the triangular lattice. For the QLM with only the kinetic term, that is relevant to the original spin model, 
we find that the ground state of the QLM displays crystalline behavior with a spontaneous breaking 
of six-fold rotation symmetry, but not of translation symmetry. We will then turn, in Sec.~\ref{sec:spin_model}, to large-scale numerical simulations of the original microscopic spin model at $m=1/6$ using direct QMC simulations. Computation of the diagonal spin structure factor confirms the existence of a stripe ordering of down spins with a moderately large correlation length. The presence of this crystalline phase is further corroborated using R\'{e}nyi entanglement entropy. Conclusions are given in Sec.~\ref{sec:conclusion}.

\section{Extended XXZ model in a magnetic field on the kagome lattice}
\label{sec:model}

\subsection{Microscopic model}

\begin{figure}
\begin{center}
\includegraphics[width=0.6\columnwidth,clip]{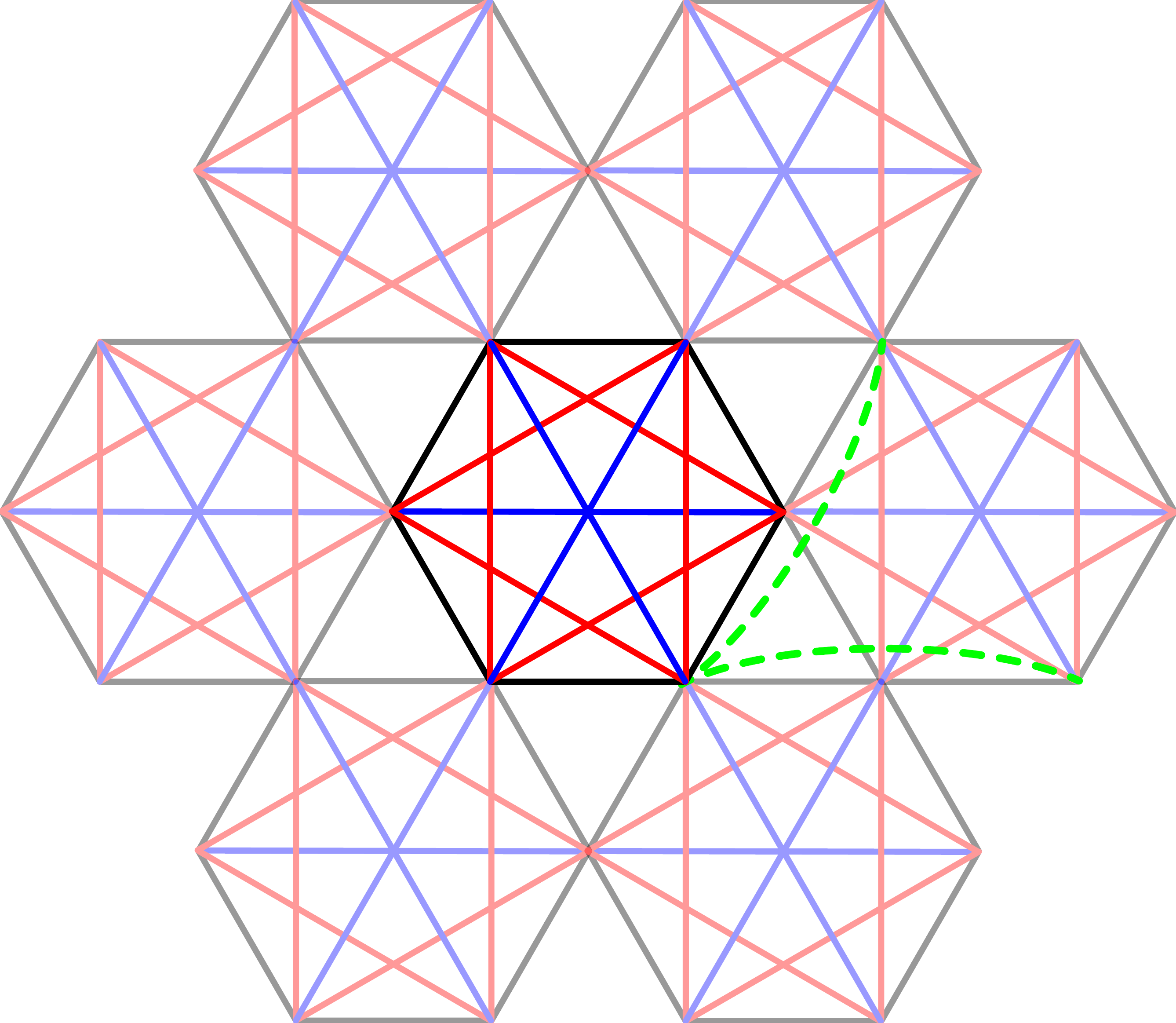}
\end{center}
\caption{(Color online) Kagom\'e lattice with the interactions between the first (black), second (red) and third (blue) neighbours within the same hexagonal plaquettes, as included in the model (\ref{eq:BFG_model}). 
Typical examples of the third-nearest-neighbor interaction between different hexagons 
[that are not included in the Hamiltonian \eqref{eq:BFG_model}] are shown by green dashed lines. 
Others are obtained by space-group operations.}
\label{fig:kagome_lattice}
\end{figure}

In the search for simple microscopic models (e.g. with two-spin interactions) hosting quantum spin
liquid phases, BFG introduced a spin-1/2 XXZ model on the Kagom\'e lattice~\cite{Balents2002} with the aim to reproduce, in a certain regime of parameters, the physics of a (generalized) quantum dimer model, known for hosting a $\mathbb{Z}_2$ topological spin liquid phase. We reproduce here their construction for completeness and obtain the effective Hamiltonians for the plateaus. The Hamiltonian is defined on the hexagonal plaquettes of the Kagom\'e lattice (see Fig.~\ref{fig:kagome_lattice}) and reads:
\begin{equation}
\begin{split}
&H = H_0 + H_{xy},\\
&H_0 = \sum_{\hex} \left[ J_{z}\sum_{\langle i,j \rangle\in \hex } S^z_i S^z_j - h\sum_{i \in \hex} S^z_i \right], \\
& H_{xy} = \sum_{\hex} J_{xy} \sum_{\langle i,j \rangle\in \hex } (
S^{+}_{i}S^{-}_{j} + S^{-}_{i}S^{+}_{j} ),
\end{split}
\label{eq:BFG_model}
\end{equation}
where ${\mathbf S}_i$ is the spin-1/2 operator. $H_0$ is the diagonal part of the Hamiltonian made up of the Ising interactions between first, second and third neighbors {\em within} each plaquette (see Fig.~\ref{fig:kagome_lattice} for all the corresponding links) as well as the Zeeman term for a magnetic field $h$ along the $z$-axis.  
Note that the third-neighbor interactions between sites belonging to different hexagons, 
shown by the green dashed lines in Fig.~\ref{fig:kagome_lattice}, are not included 
(see Sec.~\ref{sec:mapping-to-QDM} for the effect of these neglected interactions).  
The second part $H_{xy}$ contains the spin-flip terms between sites within the same hexagon. 
Originally, BFG considered all such terms on each plaquette, but, as will be clear below, setting $J_{xy}=0$ for the second and third neighbors does not alter the physical behavior [see Eq.~\eqref{eqn:generalized-Jring}]. Compared to the original Hamiltonian defined in Ref.~\onlinecite{Balents2002}, we added the magnetic field term and the coefficient of the $J_z$ terms is modified by a factor of $1/2$. The model can also be reformulated in terms of hardcore bosons, with the boson density playing the role of the magnetization. In this language, the plateaus at $m=1/6$ and $m=1/3$ that we will investigate correspond to $1/3$ and $1/6$ filling, respectively (and their respective particle-hole values). 

\begin{figure}
\begin{center}
\includegraphics[width=\columnwidth,clip,clip]{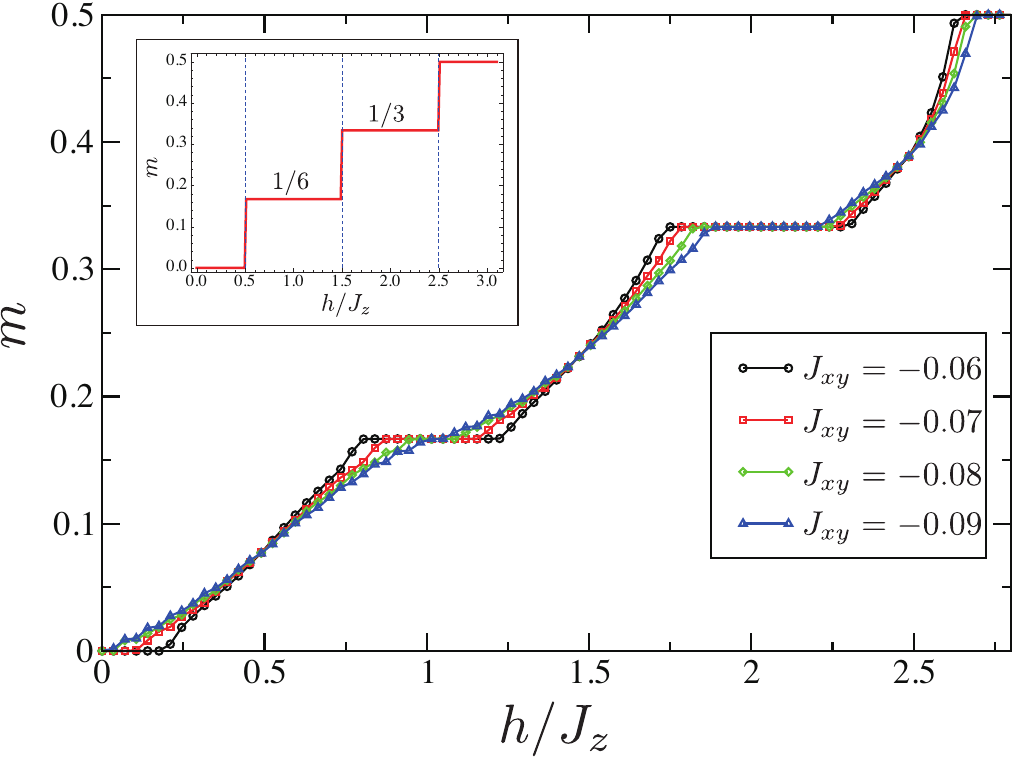}
\end{center}
\caption{(Color online) Ground state magnetization curves for a system of size $L=6$ and different values of $J_{xy}$, in the ground state ($J_z=1$ is taken as the unit of energy). Inset: Magnetization curve for $J_{z}=0$ (Ising limit). The three plateaus at magnetizations $m=0,1/6,1/3$ are clearly visible for the lowest value of the off-diagonal coupling constant  $J_{xy}=-0.06$, and gradually disappear as $|J_{xy}|$ gets larger.}
\label{fig:MofH}
\end{figure}

In the following we will be interested in the strongly anisotropic limit $J_z\gg |J_{xy}|$ where we will see that the dimer physics emerges. Since the diagonal part in \eqref{eq:BFG_model} can be written as:
\begin{equation}
H_0 = \frac{1}{2} J_z \sum_{\hex} (S^z_{\hex})^2 - h\sum_{\hex} S^z_{\hex} - \frac{3}{4} N_{\hex} J_z,
\label{eq:ice_limit}
\end{equation}
with $N_{\hex}$ the number of hexagons, every state satisfying the constraint $S^z_{\hex}=0,1,2,3$ (depending on the value of $h$) on all the plaquettes is a ground state of the classical part $H_0$. First excited states are separated by a gap of magnitude $J_z$. As a consequence, the magnetization curve $m(h)$ (with normalization $m=\sum_i S^z_i / N$) displays plateaus of width $J_z$ at the values $m=0,1/6,1/3,$ (see the inset of Fig.~\ref{fig:MofH}) and saturation at $1/2$ 
($m=(3-n_{\downarrow})/6$ where $n_{\downarrow}=3,2,1,0$ is the number of down spins per hexagon) 
which are expected to survive for a finite value of $J_{xy}$. As an illustration, we present in Fig.~\ref{fig:MofH} the magnetization curve in the ground state for a moderate system size $L=6$ and finite small values of $J_{xy}$, as obtained from QMC simulations (see Sec.~\ref{sec:spin_model} for details and a finite-size scaling analysis of the size of the $m=1/6$ plateau).

The above construction leads to magnetization plateaus where the ground states degeneracies scale exponentially with the system size~\cite{Balents2002}. Dynamics induced by the off-diagonal terms will lift this degeneracy, as presented below. 

\subsection{Mapping to generalized quantum dimer models}
\label{sec:mapping-to-QDM}
\begin{figure}
\begin{center}
\includegraphics[width=\columnwidth,clip]{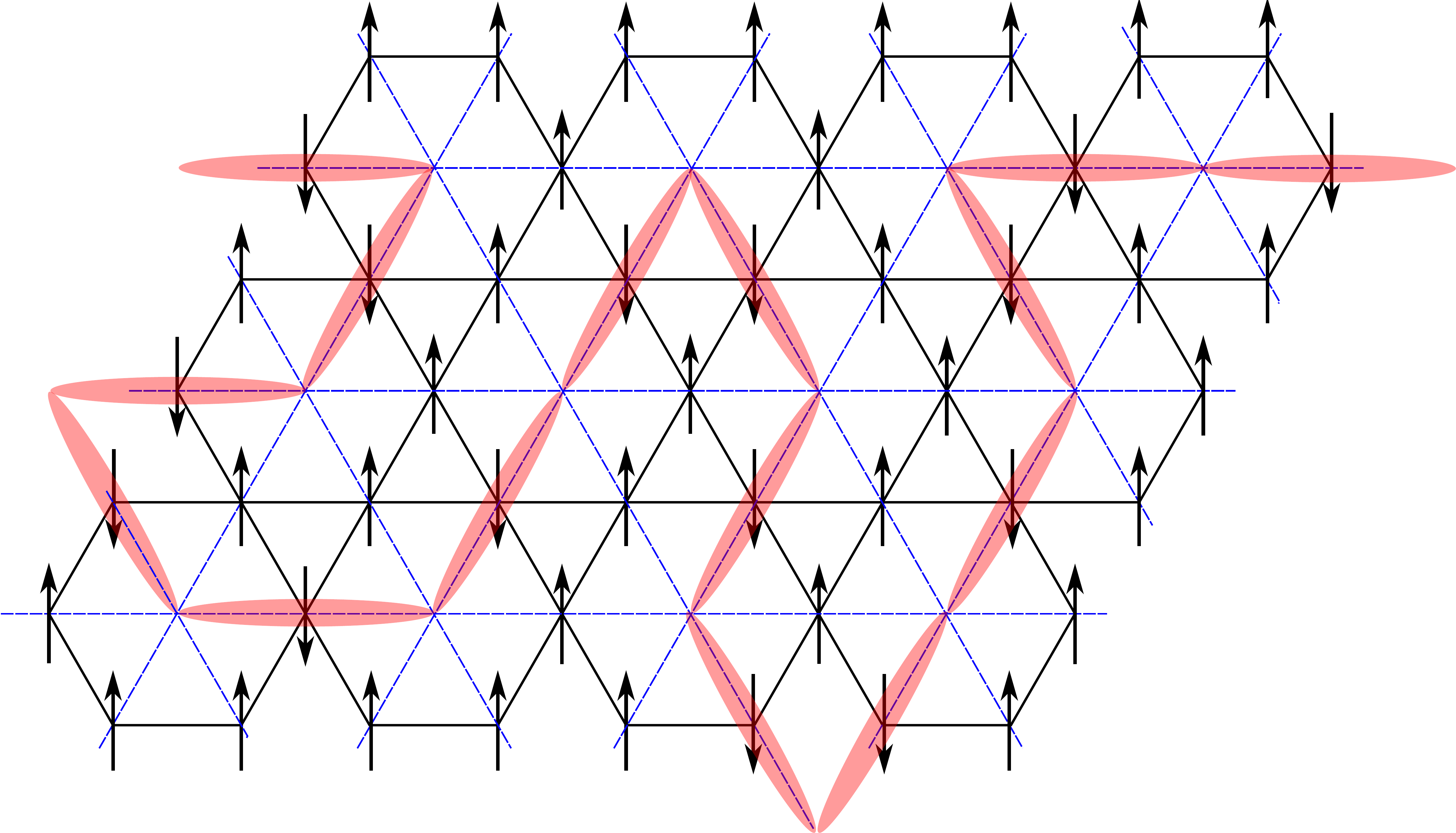}
\end{center}
\caption{(Color online) Representation of one of the ground states of model \eqref{eq:BFG_model} for $J_{xy}=0$ and $m=1/6$ in terms of dimers (red) on the underlying triangular lattice (blue). Loop configurations are formed due to the constraint of two dimers per triangular lattice site.}
\label{fig:dimer_mapping}
\end{figure}

A simple way to visualize the local constraints $S^z_{\hex}=0,1,2,3$ (for all hexagons) in the Ising limit is to faithfully map the ground state configurations to those of dimers: draw a dimer on a link of the triangular lattice formed by the centers of hexagons when the middle of the link is occupied by a down spin, and leave the same link empty for an up spin. The magnetization plateaus $m=(3-n_{\downarrow})/6$ are then reproduced by having a constraint such that every triangular lattice site must be touched by $n_{\downarrow}$ dimers. This procedure is illustrated in Fig.~\ref{fig:dimer_mapping} for the $m=1/6$ ($n_{\downarrow}=2$) plateau with two dimers per site, which allows us to visualize the ground state spin configurations of the diagonal part $H_{0}$ as fully-packed self-avoiding loops on the triangular lattice.

\begin{figure}
\begin{center}
\includegraphics[width=\columnwidth,clip]{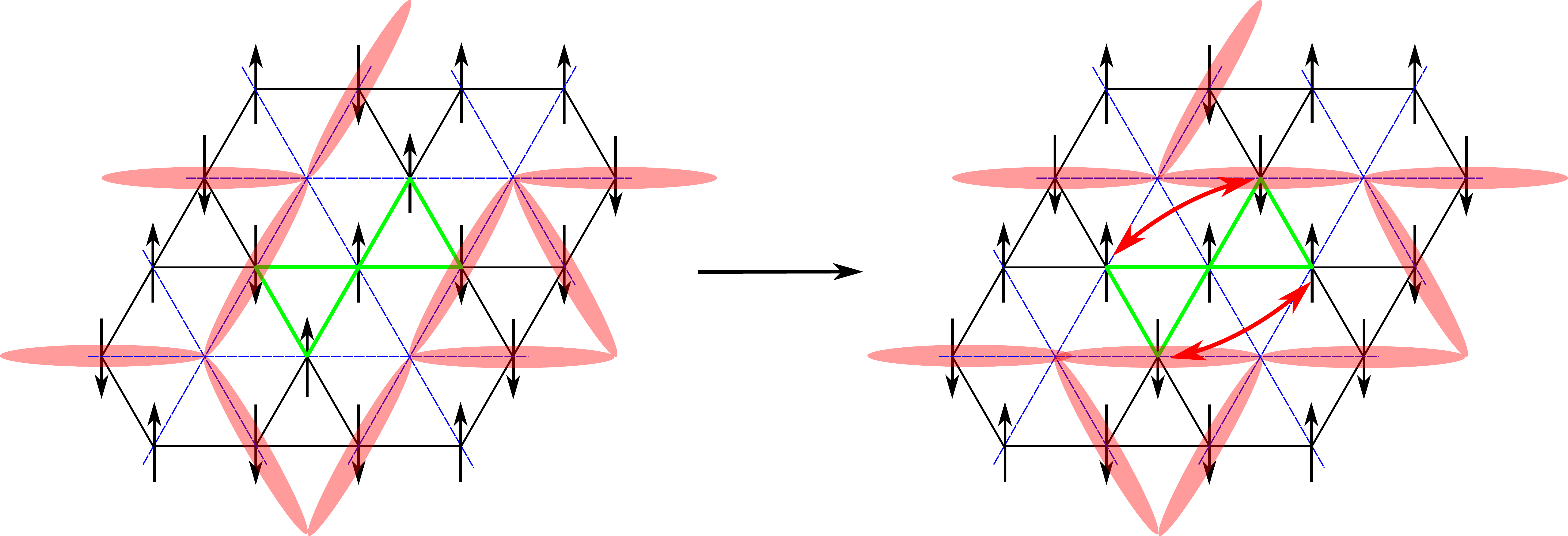}
\end{center}
\caption{(Color online) Off-diagonal second-order process of $H_{xy}$ on a flippable bow-tie plaquette (green) for $n_{\downarrow}=3$. The antiparallel spins of two flippable pairs are exchanged, corresponding to a flip of two parallel dimers on the underlying triangular lattice (blue).}
\label{fig:dimer_hopping}
\end{figure}

The generalized quantum dimer models appear when treating $H_{xy}$ in degenerate perturbation theory. Clearly, first-order processes do not contribute, since applying the spin-flip term only once produces configurations with two excited hexagons, thus outside of the ground state manifold. In second-order, two types of processes are allowed. The first process simply flips the same pair of spins twice, merely shifting the total energy by $\Delta E^{(2)}_{\mathrm{diag}}=-9(1-m^2)N_{\hex}J_{xy}^2/J_z$. The second process involves a flippable bow-tie plaquette, as pictured in green in Fig.~\ref{fig:dimer_hopping}. The two pairs of antiparallel spins are flipped, generating another state fulfilling the ground state constraint. The second-order effective Hamiltonian is thus given by the ring-exchange Hamiltonian~\cite{Balents2002}:
\begin{equation}
H_{\mathrm{eff}}^{(2)} = -J_{\mathrm{ring}} \sum_{\bowtie} ( S^{+}_1 S^{-}_2 S^{+}_3 S^{-}_4 + \mathrm{h.c.} ),
\label{eq:ring_hamiltonian}
\end{equation}
where the sum runs over all flippable bow-tie plaquettes (labels $1,2,3$ and $4$ denote the exterior sites of the bow-ties) 
and the ring-exchange amplitude is given by $J_{\mathrm{ring}}=4J_{xy}^2/J_z$. Quite importantly, the sign of $J_{xy}$ is immaterial which allows numerical simulations of the original microscopic Hamiltonian (\ref{eq:BFG_model}) with QMC techniques, by considering the model with a ferromagnetic $J_{xy}<0$ coupling.  
Also in order to obtain the ring-exchange Hamiltonian \eqref{eq:ring_hamiltonian}, only the XY-interactions 
on the nearest-neighbor ($J_{\text{xy}}^{\text{NN}}$) and the next-nearest-neighbor ($J_{\text{xy}}^{\text{NNN}}$)  
bonds are crucial:
\begin{equation}
J_{\text{ring}} = \frac{2\{(J_{xy}^{\text{NN}})^{2}+(J_{xy}^{\text{NNN}})^{2}\}}{J_{z}} \; .
\label{eqn:generalized-Jring}
\end{equation}
From this, one can see, as stated previously, that suppressing the further-neighbor XY interactions in $H_{xy}$ [Eq.~\eqref{eq:BFG_model}] does not alter physics qualitatively and simply rescales $J_{\mathrm{ring}}$ by a factor 1/2.
We exploit this property in Sec.~\ref{sec:spin_model}.   

In the dimer language, the ring move on a bow-tie of the original lattice corresponds to the flip of two parallel dimers on a diamond of the dual triangular lattice (Fig.~\ref{fig:dimer_hopping}). This is nothing but the off-diagonal term of the quantum dimer model~\cite{Rokhsar1988} on a triangular lattice~\cite{Moessner2001}, which, as usual, is conveniently extended with a diagonal potential ($V$) term which counts the number of flippable plaquettes (diamonds):
\begin{equation}
\begin{split}
H_{\mathrm{eff}}^{(2)}= & -t \sum_{{\includegraphics[scale=0.25]{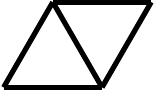}}} \left( | \raisebox{-0.65ex}{\includegraphics[scale=0.45]{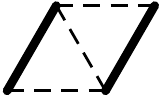}} \rangle  \langle \raisebox{-0.65ex}{\includegraphics[scale=0.45]{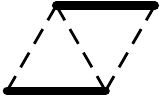}} |
+ | \raisebox{-0.65ex}{\includegraphics[scale=0.45]{dimer2}} \rangle  \langle \raisebox{-0.65ex}{\includegraphics[scale=0.45]{dimer1}} | \right) \\
& + V \sum_{{\includegraphics[scale=0.25]{plaq}}} \left( | \raisebox{-0.65ex}{\includegraphics[scale=0.45]{dimer1}} \rangle  \langle \raisebox{-0.65ex}{\includegraphics[scale=0.45]{dimer1}} |
+ | \raisebox{-0.65ex}{\includegraphics[scale=0.45]{dimer2}} \rangle  \langle \raisebox{-0.65ex}{\includegraphics[scale=0.45]{dimer2}} | \right).
\end{split}
\label{eq:QDM_triangular}
\end{equation}
In the Ising limit $J_{z}\gg J_{xy}$, low-energy physics of the BFG model is expected to be described by this Hamiltonian for $V=0$.  
The following four-spin interaction \cite{Balents2002,Sheng2005}
\begin{equation}
\begin{split}
& Q_4 \sum_{\bowtie} \left\{ 
\left(1/2 -S_{1}^{z} \right)\left(1/2 + S_{2}^{z} \right)\left(1/2 - S_{3}^{z}\right)\left(1/2 + S_{4}^{z} \right)  \right. \\
& + \left. 
\left(1/2 + S_{1}^{z}\right)\left(1/2 - S_{2}^{z} \right)\left(1/2 + S_{3}^{z} \right)\left(1/2 - S_{4}^{z} \right)  
\right\} 
\end{split}
\label{eqn:4-spin-int}
\end{equation}
($1,2,3,4$ label the four spins on a bow-tie in a anti-clockwise way) 
reduces, in the same limit, to the $V$-term ($V=Q$) as it counts the number of 
flippable plaquettes.   
For the $m=1/3$ plateau corresponding to one dimer (or, one $\downarrow$ spin) per hexagon,  
one may use instead the {\em inter-hexagon} (two-spin) third-neighbor interaction 
\begin{equation}
J_{z}^{(3)}\sum_{\substack{%
\langle i, j\rangle\\
\text{inter-hexagon}}} S_{i}^{z}S_{j}^{z} 
\label{eqn:inter-hex-3rdN}
\end{equation}
that is not contained in model \eqref{eq:BFG_model} to realize the dimer interaction $V=J_{z}^{(3)}$.  

Thus, one sees that {\em the effective model is identical for the three magnetization plateaux, the only difference coming from the constraint on the Hilbert space}. For the constraint of one dimer per site (corresponding to the plateau $m=1/3$), we have the usual QDM~\cite{Moessner2001}, while in the case with two dimers per site (the plateau at $m=1/6$), a QLM is obtained as announced in the introduction. The original model derived by BFG \cite{Balents2002} describing the $m=0$ plateau corresponds to three dimers per site. Table~\ref{tab:effective-model-summary} summarizes the nature of the effective models for the different plateaus.

The mapping to dimer models [including the diagonal $V$-term in Eq.~\eqref{eq:QDM_triangular}] allows us to benefit from the accumulated knowledge on this family of models~\cite{Rokhsar1988,Moessner2001,Ioselevich2002,Misguich2002,Ralko2005,Syljuasen2005,Ralko2006,Vernay2006,Ralko2007,Misguich2008a,Herdman-W-11}. QDMs generically admit a Rokhsar-Kivelson~\cite{Rokhsar1988} (RK) point $V=t$, where the ground-state wave-function is exactly known. For the QDM on the triangular lattice, the ground state is exactly shown to be a topological gapped $\mathbb{Z}_2$ liquid phase~\cite{Moessner2001,Fendley2002,Ioselevich2002}, which furthermore extends in the region $V/t<1$.
\begin{figure*}[t]
\begin{center}
\includegraphics[width=0.9\textwidth,clip]{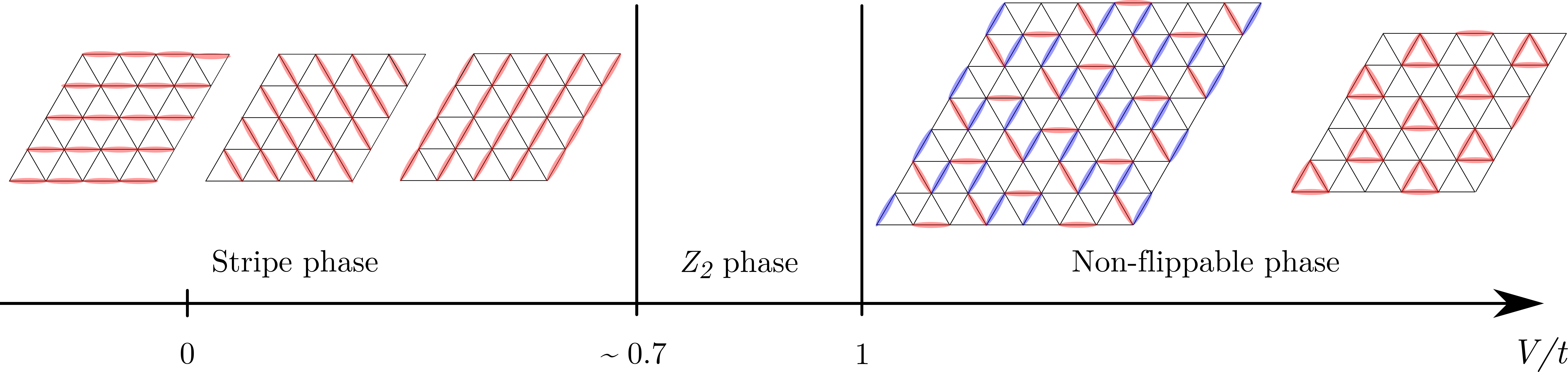}
\end{center}
\caption{(Color online) Phase diagram of the QLM Eq.~(\ref{eq:QDM_triangular}) as a function of $V/t$, containing the three different phases discussed in Sec.~\ref{sec:loop_general}. The boundary between stripe phase and 
$\mathbb{Z}_{2}$ spin liquid is determined using topological degeneracy and stripe structure factor.}
\label{fig:loop_phase_diag}
\end{figure*}
In the three-dimer model, BFG confirmed the presence of a topological phase at the RK point~\cite{Balents2002}, showing in particular that visons are gapped, implying the presence of
deconfined fractionalized spinons~\cite{Senthil2000}.  The authors of Ref.~\onlinecite{Balents2002} also speculated that the $\mathbb{Z}_2$ liquid phase could extend beyond the $V=0$ point corresponding to
the original microscopic spin model, in contrast to the one-dimer case~\cite{Moessner2001,Ralko2006} for which the liquid phase ends at $(V/t)_{\text{c}} \simeq 0.8$~\cite{Ralko2006}. An exact diagonalization study~\cite{Sheng2005} of the Hamiltonian (\ref{eq:ring_hamiltonian}) shows the presence of a vison gap in a regime including the BFG point ($V=0$), 
thereby supporting the above suggestion. However, the presence of the spin liquid was only firmly established using QMC simulations on large systems. First, a complete numerical phase diagram of the model (\ref{eq:BFG_model}) with ferromagnetic $J_{xy}$ and $h=0$ was obtained~\cite{Isakov2006}:  at strongly negative $J_{xy}/J_z$, a planar ({\em i.e.}, superfluid in the bosonic language) phase accompanied by the breaking of the U(1) symmetry is found, as expected. When the magnitude of the spin-flip term was increased, a continuous phase transition to an apparently featureless insulating phase was observed. The phase diagram was also extended to finite temperature~\cite{Isakov2007}. Finally, a direct evidence of the topological nature of the insulator is obtained by computing the topological entanglement (R\'enyi) entropy~\cite{Isakov2011}, which was evaluated as $\log 2$ as expected for a $\mathbb{Z}_2$ spin liquid phase. As a consequence of the condensation of fractionalized excitations, the transition between the planar phase and the featureless insulator is an exotic 3D $\mathrm{XY}^{*}$ quantum critical point~\cite{Grover2010,Isakov2012}.

While the nature of the phases encountered at the $m=0$ and $m=1/3$ plateaus is well understood 
from this previous set of study, the ground state physics on the $m=1/6$ plateau has never been investigated  to our best knowledge. In the following, we focus on this situation by performing numerical simulations first on the effective QLM (Sec.~\ref{sec:loop_model}), and then on the microscopic spin model (Sec.~\ref{sec:spin_model}).

\begin{table}
\caption{\label{tab:effective-model-summary} Three plateaus of $S=1/2$ BFG model and the corresponding effective models.}
\begin{ruledtabular}
\begin{tabular}{lccc}
plateaus &   $q(S-m)$ & number of dimers/site & effective model \\
\hline
$m=0$ & $3/2$ & three & generalized QDM \\
$m=1/6$ & 
$1$ & two & QLM \\
$m=1/3$ & 
$1/2$ & one & QDM
\end{tabular}
\end{ruledtabular}
\end{table}

\section{Phase diagram of the quantum loop model}
\label{sec:loop_model}

We consider here the effective model (\ref{eq:QDM_triangular}) in the case of two dimers per site,  
where allowed states are represented by configurations of self-avoiding loops such as represented in Fig.~\ref{fig:dimer_mapping}. Motivated by the original spin model Eq.~\eqref{eq:BFG_model} at the $m=1/6$ plateau, we are primarily interested in the nature of the ground state of the effective QLM at $V/t=0$, however we will also investigate the nature of the surrounding phases in the phase diagram. We first give, in Sec.~\ref{sec:loop_general}, general arguments on the structure of the phase diagram as well as on the topological properties of the loop configurations, before complementing this analysis in Sec.~\ref{sec:loop_qmc} with QMC simulations of the loop model. Anticipating the results obtained, we present in Fig.~\ref{fig:loop_phase_diag} the ground-state phase diagram of the QLM on the triangular lattice to illustrate the following discussion.

\subsection{General considerations of the phase diagram of the quantum loop model}
\label{sec:loop_general}
When $V/t\rightarrow -\infty$, the ground-state energy is minimized by configurations where loops form straight lines 
along one of the three lattice directions, as shown in the left part of the phase diagram presented in Fig.~\ref{fig:loop_phase_diag}. As aligned up and down spins form alternating stripes in the corresponding configurations of the original spin models (see the right panel of Fig.~\ref{fig:QMCcorr}), we call this a {\em stripe phase}. In this gapped ordered phase, some of the rotations and reflections of the triangular lattice are spontaneously broken ({\em i.e.}, the point group changes from $C_{6\text{v}}$ to $C_{2\text{v}}$), leading to the three-fold degenerate ground states 
(in the QDM, the corresponding columnar phase further breaks translation symmetry).

At the RK point $V/t=1$, the ground state is given by the equal-weighted sum of all fully-packed loop configurations on the triangular lattice. To our best knowledge, the equivalent classical problem has not been studied before. As the triangular lattice is non-bipartite, we expect that the loop segment correlations are short-ranged, indicating a {\it liquid} phase presumably with a gap, as also found for the RK point of the QDM~\cite{Moessner2001}.

\begin{figure}
\begin{center}
\includegraphics[width=0.9\columnwidth,clip]{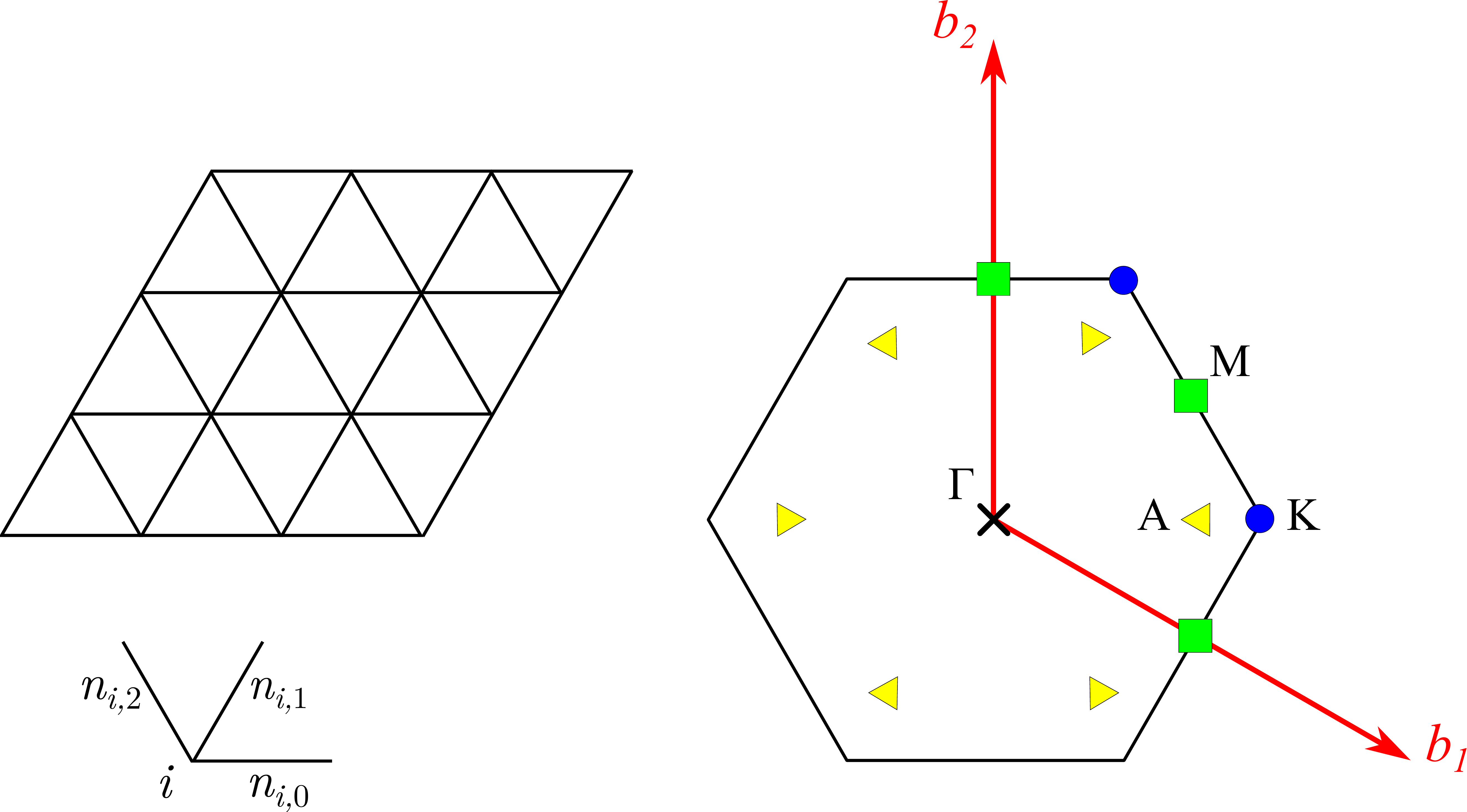}
\end{center}
\caption{(Color online) Top left: Triangular clusters used in the QLM study. Bottom left: notations for the loop segment occupation number. Right: First Brillouin zone of the triangular lattice, with the reciprocal space vectors $\mathbf{b}_1$ and $\mathbf{b}_2$. The high symmetry points $K=(4\pi/3,0)$, $M=(\pi,\pi/\sqrt{3})$ and $A=(\pi,0)$ required for the non-flippable states are represented.}
\label{fig:brillouin_zone}
\end{figure}

When $V/t>1$, the ground states are readily found to have zero energy and correspond to non-flippable loop configurations ({\it i.e.} for which the kinetic term vanishes), which is again similar to the staggered phase of the QDM.  
While the construction of those states is quite straightforward in the QDM, we have found that for $L \times L$ systems, the QLM accepts two different families of non-flippable configurations, both of which are shown in the right part of the phase diagram of Fig.~\ref{fig:loop_phase_diag}. The first family is obtained by starting from the staggered states of the QDM \cite{Moessner2001,Ralko2005}, which are 12-fold degenerate. On the left, one of those 12 states is represented in red, with only two of the three types of links (corresponding to the three lattice directions 0, 1, and 2 in Fig.~\ref{fig:brillouin_zone}) occupied. A non-flippable loop configuration can then be generated by adding parallel (blue) dimers on half of the remaining type of links, for which there are two possibilities (a one-step translation in the direction of these parallel dimers also creates a different non-flippable state). The same prescription can be applied starting from the other QDM staggered states, leading to a total degeneracy of $2\times 12=24$. The second family is obtained by tiling the lattice with the shortest possible triangular loops, as depicted on the right part of Fig.~\ref{fig:loop_phase_diag}. The degeneracy of this family is readily found to be 6. Considering the Brillouin zone drawn in the right panel of Fig.~\ref{fig:brillouin_zone}, the first family of non-flippable states requires the cluster to have both $A$ and $M$ points (i.e. $L$ multiple of 4), while the second one needs the $K$ point (i.e. $L$ multiple of 3). We finally remark that the nature of non-flippable states (or absence thereof) can be different for other types of clusters (rectangular clusters for instance).

From these considerations, the point $V=0$ lies between the two limiting cases $V/t\rightarrow - \infty$ (stripe) and $V/t \rightarrow 1^{-}$ (liquid). As both the stripe and liquid phases are gapped, they should extend in a finite region of phase space close to these limits, and it is not clear {\it a priori} in which phase the point $V=0$ is located. Of course, it is also possible that one or several intervening crystalline phases arise: natural candidates are equivalent of plaquette, or $\sqrt{12}\times \sqrt{12}$ phases observed in QDMs~\cite{Moessner2001,Ralko2005,Ralko2006,Syljuasen06}. This can only be determined by exact numerical calculations, which will be presented below in Sec.~\ref{sec:loop_qmc}.

\begin{figure}
\begin{center}
\includegraphics[width=0.8\columnwidth,clip]{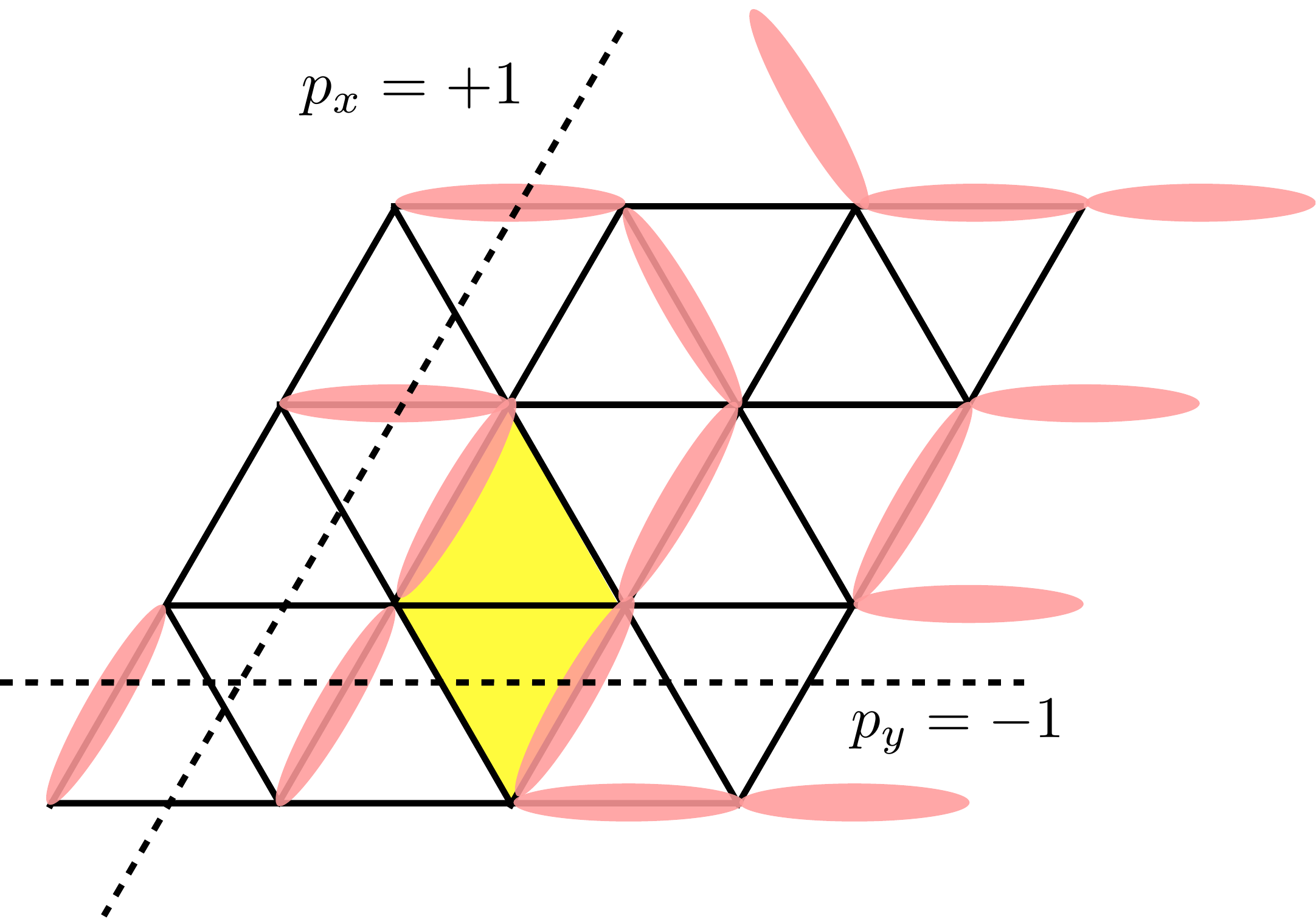}
\end{center}
\caption{(Color online) Drawing two lines on the triangular lattice
  allows to define two conserved $\mathbb{Z}_2$ invariants $(p_x,p_y)$ associated
  to the parity of the number of loop segments crossed by the line. In
  this example, the number of loop segments pointing in the vertical $y$ (horizontal $x$) direction which cross the line is odd (even), corresponding to $p_y=-1$ ($p_x=1$). This number does not depend on
  the precise location of the line (parallel lines shifted by one or
  several lattice units result in the same parity). The
  flipping terms of the Hamiltonian, acting for instance on
  the shaded plaquette (other flips are possible in this
  configuration), do not change these two parities. Note that a
  loop segment located on a diagonal bond of a plaquette contributes
  to both $p_x$ and $p_y$.}
\label{fig:loop_winding}
\end{figure}

It is interesting to note at this stage that on a manifold with non-zero genus (cylinder, torus, etc), the QLM model possesses topological sectors which can be defined in the same way as in the QDM case. As illustrated in Fig.~\ref{fig:loop_winding} for a torus, the parity of the number of loop segments crossing a dashed line in either the $x$ or $y$ directions of the torus is conserved by the dynamics of the Hamiltonian.  
On the torus, this defines a pair of $\mathbb{Z}_2$ invariants $p_x=\pm 1$ and $p_y=\pm 1$, and thus divides the full Hilbert space into four distinct topological sectors, which we label by $(p_x,p_y)$. The existence of these four topological sectors is the key to identify the $\mathbb{Z}_2$ spin liquid phase close to $V/t=1$, as {\it e.g.} in the QDM on the triangular lattice~\cite{Moessner2001}.
One may wonder if the $\mathbb{Z}_2$ spin liquid found here is the same as that 
in the usual triangular-lattice QDM since even (odd) number of dimers exiting 
from each site in the QLM (QDM) suggests a mapping to an even (odd) Ising gauge theory~\cite{Moessner01b}.  
However, this difference is not important in considering the underlying nature of the spin liquids.  
In fact, by explicitly constructing the ground state of the toric code~\cite{Kitaev-03} on a triangular lattice~\footnote{%
Here the plaquette operator is defined on each elementary triangle and the star operator on six bonds emanating 
from each site.}, one can show that both spin liquids share the same quasiparticle (anyon) contents and exhibit  
the same ground-state degeneracy originating from the $\mathbb{Z}_{2}$ flux, which we will use in the following.  

In the topological $\mathbb{Z}_{2}$ spin liquid phase, the four topological sectors (on a torus) 
corresponds to the four different ways of threading the $\mathbb{Z}_{2}$ flux through the two periods 
of the system that cannot be eliminated by finite-order perturbations \cite{Wen1991}. Therefore, the four sectors, in the topological spin liquid phase, are separated from each other by exponentially small gaps ({\em topological gaps}) and get degenerate in the thermodynamic limit. As was previously observed for the QDM~\cite{Ralko2005}, we expect the liquid phase to be detected by the closing of the above topological gaps.  

On the other hand, in the ordered stripe phase, topological sectors which do not contain the pure stripe 
configurations (shown in the left part of Fig.~\ref{fig:loop_phase_diag}) cannot optimize 
the attractive $V$-term and should have a higher energy: correspondingly, the topological gap should be finite even in the thermodynamic limit, providing us with a practical method to distinguish the two phases. One should however be careful in identifying the sectors to which the stripe configurations belong. Indeed on a torus with $L_x=L_y=L$, the QLM can be studied for even and odd $L$ (in contrast to the QDM that can only host even-$L$ samples). When $L$ is odd, the three stripe configurations shown in Fig.~\ref{fig:loop_phase_diag} are located in the three different sectors $(-1,1)$, $(1,-1)$ and $(-1,-1)$ which are strictly identical (the number of configurations is the same and the corresponding blocks of the Hamiltonian are identical): we define the topological gap for odd-$L$ samples as $\Delta_{\text{T}}^{\text{odd}}=E_0(1,1)-E_0(-1,-1)$, where $E_0(p_x,p_y)$ is the ground state energy in the $(p_x,p_y)$ sector. For even $L$, the three stripe configurations belong to the $(1,1)$ sector, and all other sectors are identical, and we use $\Delta_{\text{T}}^{\text{even}}=E_0(-1,-1)-E_0(1,1)$.

\subsection{Phase diagram: QMC results} 
\label{sec:loop_qmc}
We now turn to QMC simulations of the QLM on a torus (for clusters with the geometry represented in Fig.~\ref{fig:brillouin_zone}, with $L_x=L_y=L$ and $N=L^2$ sites), as defined in Eq.~\eqref{eq:QDM_triangular}. We use the reptation QMC~\cite{Baroni01}, which projects a given initial configuration belonging to each topological sector onto the ground state in the same sector. We use a formulation of reptation QMC similar to the one presented in Ref.~\onlinecite{Syljuasen06}: to accelerate the convergence, we work in a continuous-time formalism as well as with a guiding wave function that
provides a finite fugacity for each flippable plaquette of the triangular lattice (an optimized value for the fugacity is obtained from a short preliminary run). We simulated even-$L$ samples up to $L=10$ (and one data point at $L=12$), and odd-$L$ samples up to $L=9$. 

\begin{figure}
\begin{center}
\includegraphics[width=\columnwidth,clip]{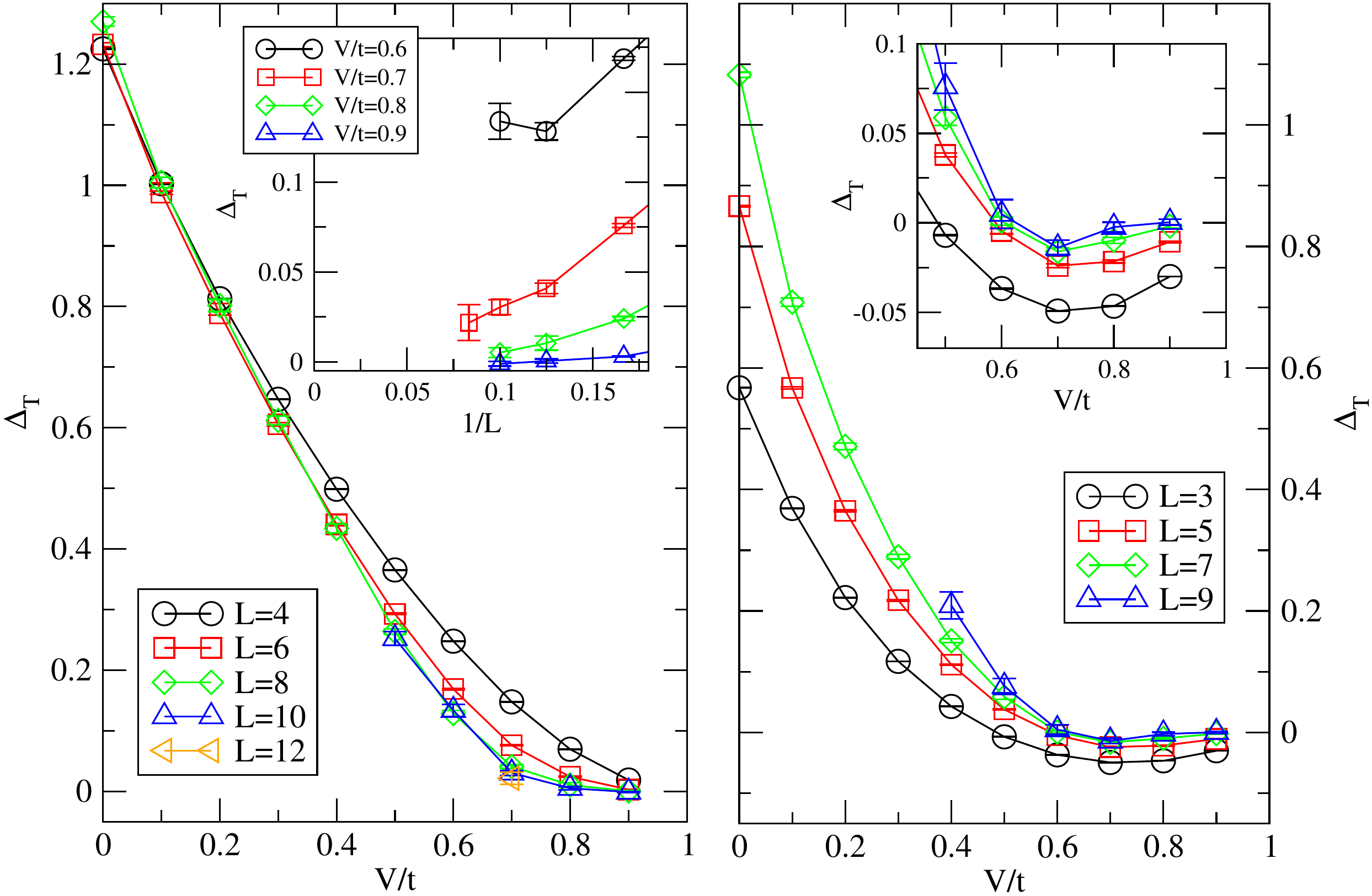}
\end{center}
\caption{(Color online) Topological gap $\Delta_{\text{T}}$ as a function of $V/t$, for different sample sizes $L$.  
The finite-size effects are different for even ($\Delta^{\text{even}}_{\text{T}}$; left panel) 
and odd ($\Delta^{\text{odd}}_{\text{T}}$; right panel) 
values of $L$.  In the thermodynamic limit, we expect that in both cases, $\Delta_{\text{T}}$ if finite 
for the stripe phase $V/t<(V/t)_{\text{c}}$, and vanishes for the $\mathbb{Z}_{2}$ spin liquid phase 
$(V/t)_{\text{c}} \leq V \leq 1$. Left inset: Finite-size scaling as a function of $1/L$ for the even-$L$ samples provides an estimate $(V/t)_{\text{c}} \simeq 0.70(5)$ of the value where the gap first vanishes. Right inset: Zoom close to the RK point for odd-$L$ samples.}
\label{fig:loop_gap}
\end{figure}

We first present the results for the topological gap $\Delta_{\text{T}}$, as obtained by computing the ground state energy in each non-equivalent sector using the standard mixed estimator~\cite{Baroni01}. Fig.~\ref{fig:loop_gap} shows $\Delta_{\text{T}}$ as a function of $V/t$ for even (left panel) and odd (right panel) $L$, for different system sizes. For even samples, the topological gap $\Delta^{\text{even}}_{\text{T}}$ has a monotonous behavior: it is clearly finite for low values of $V/t$ and vanishes with system size close enough to the RK point $V/t=1$.  
Finite-size scaling as a function of $1/L$ (left inset of Fig.~\ref{fig:loop_gap}) indicates that the gap vanishes around $(V/t)_{\text{c}} \simeq 0.70(5)$, separating the stripe phase for $V/t <(V/t)_{\text{c}}$ from the topological $\mathbb{Z}_{2}$ liquid phase for $(V/t)_{\text{c}} <V/t\leq 1$. For the largest samples, our data at $V/t=0.8$ and $V/t=0.9$ show that the ground state energies in different topological sectors are identical within error bars (we cannot resolve the exponentially small splitting in the liquid phase).

The gap for the odd-$L$ samples displays an apparently different behavior, with the topological sector $(1,1)$ having a higher energy for low enough $V/t$, resulting in $\Delta^{\text{odd}}_{\text{T}}>0$ as the $V$-term favors the sectors $(-1,1)$, $(1,-1)$ and $(-1,-1)$. On the other hand, when the system is close enough to the RK point, the sector
$(1,1)$ has lower energy (resulting in $\Delta^{\text{odd}}_{\text{T}}<0$ in our definition
of $\Delta^{\text{odd}}_{\text{T}}$) for finite-size samples. If we assume that the ordering of the four levels in the spin liquid phase is the same as that for the even-$L$ samples, the sign change marks the transition. The value of $V/t$ at which $\Delta^{\text{odd}}_{\text{T}}=0$ varies with system size and appears to extrapolate to a consistent value for the critical point $(V/t)_{\text{c}} \simeq 0.70(5)$ (see right inset of Fig.~\ref{fig:loop_gap}). Here again, for the largest odd-$L$ samples, the energies in all topological sectors are indistinguishable within error bars for $V/t=0.8$ and $V/t=0.9$, indicating the topological degeneracy.

The topological gap for both odd and even sample is clearly finite for $V/t=0$ (with a value
$\simeq 1.3 t$, as estimated from the even-$L$ data, for which there is almost no size dependence), which indicates that this point is located in an ordered phase (presumably, in the stripe phase). In order to confirm this, we computed the loop segment occupations (using middle-slices in the reptile representation~\cite{Baroni01}) in the topological sector $(1,1)$ (which hosts all three stripe configurations) for the even-$L$ samples. This allows to determine the stripe structure factor, which we define as:
\begin{equation}
\begin{split}
S_{\rm stripe}  & =  \frac{1}{N} \sum_{\substack{ i,j \\ \alpha,\beta=0,1,2}} \langle n_{i,\alpha} n_{j,\beta} \rangle e^{\frac{2 i \pi}{3}(\beta-\alpha)} \\
& =  \frac{1}{N} \left(  \sum_{\alpha=0,1,2} \langle N_\alpha^2 \rangle - \sum_{\alpha \neq \beta} \langle N_\alpha N_\beta \rangle \right) .
\end{split}
\label{eqn:stripe-str-factor}
\end{equation}
where $n_{i,\alpha}=1$ if there is a loop segment at site $i$ occupying a link in the direction $\alpha$($=0,1,2$; the three values taken by $\alpha$ correspond to the three different directions as represented in the left bottom part of Fig.~\ref{fig:brillouin_zone}) and $n_{i,\alpha}=0$ otherwise.    
In the second line of Eq.~\eqref{eqn:stripe-str-factor}, we have introduced the notation $N_\alpha=\sum_i n_\alpha$ for the total number of bonds in the direction $\alpha$. The structure factor divided by the sample size is expected to converge to the square of the stripe order parameter $m_{\text{stripe}}$ in the thermodynamic limit: $S_{\text{stripe}}/N \rightarrow m_{\text{stripe}}^2$, with a non-zero $m_{\rm stripe}$ if the ground state exhibits the long-range stripe order. 

\begin{figure}[h]
\begin{center}
\includegraphics[width=\columnwidth,clip]{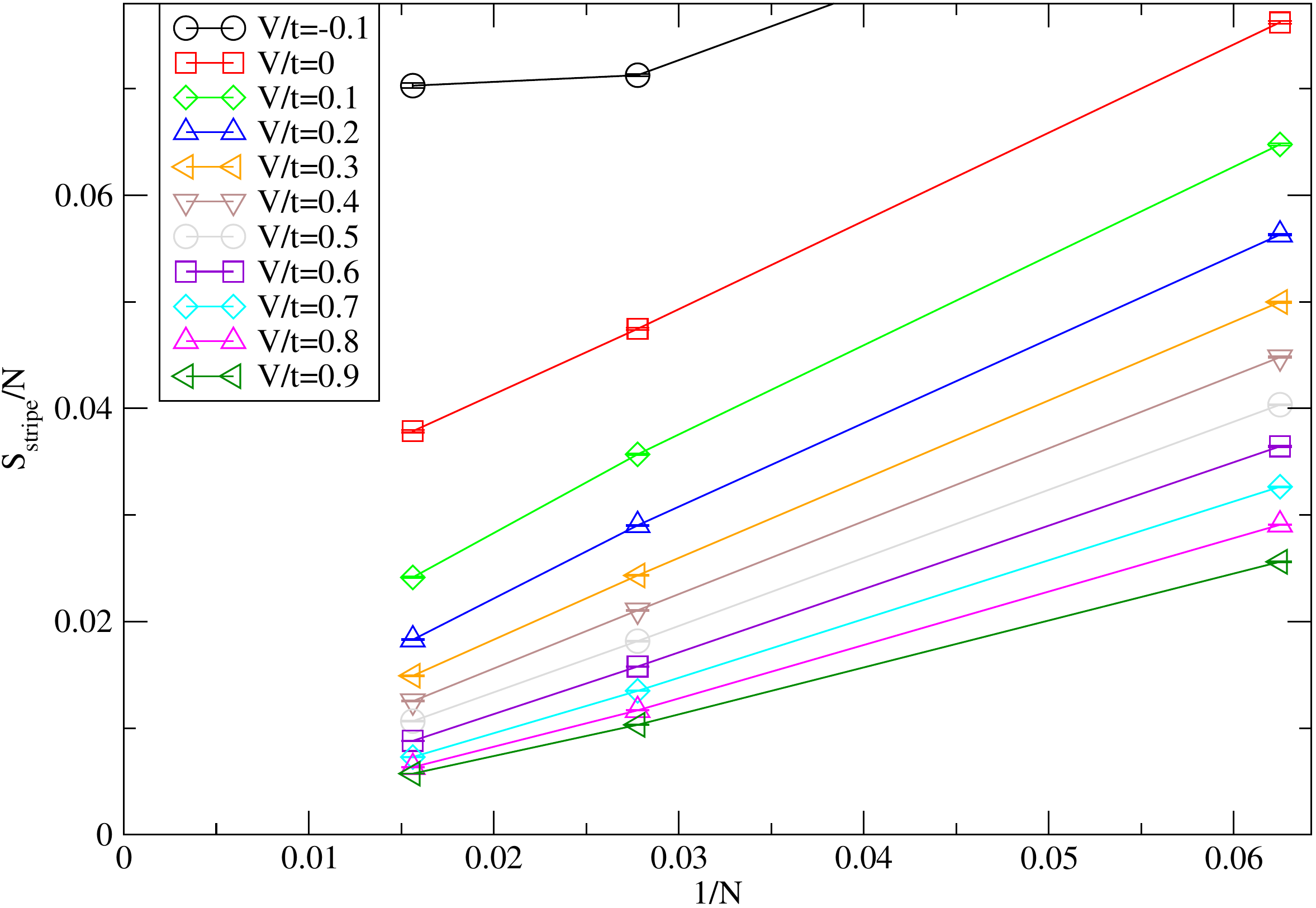}
\end{center}
\caption{(Color online) The stripe structure factor $S_{\rm stripe}$ as a function of inverse system size, for different values of $V/t$.}
\label{fig:loop_sf}
\end{figure}

The stripe structure factor $S_{\text{stripe}}/N$ is shown in Fig.~\ref{fig:loop_sf} for different values 
of the coupling $V/t$ as a function of inverse size $1/N$ (for $L=4,6,8$). We expect that $S_{\text{stripe}}/N$ should saturate in an exponential way (due to the finite correlation length) to $m_{\rm stripe}^2>0$ in the stripe phase, and to $0$ in a phase with no stripe order (at the critical point, it should decay to $0$ with a non-trivial power). Due to the moderate system sizes accessible, we cannot expect to see the exponential decay: the extrapolation to the thermodynamic limit is indeed delicate for large values of $V/t$. Considering the data for $V/t \lesssim 0.6$ (see Fig.~\ref{fig:loop_sf}), we observe that the structure factor extrapolates to a non-zero value. From this data set, we obtained an estimate for the critical point $(V/t)_{\text{c}}=0.65(15)$, which is less precise, though consistent, than the one determined above by the topological gap. 
\begin{figure}
\begin{center}
\includegraphics[width=\columnwidth,clip]{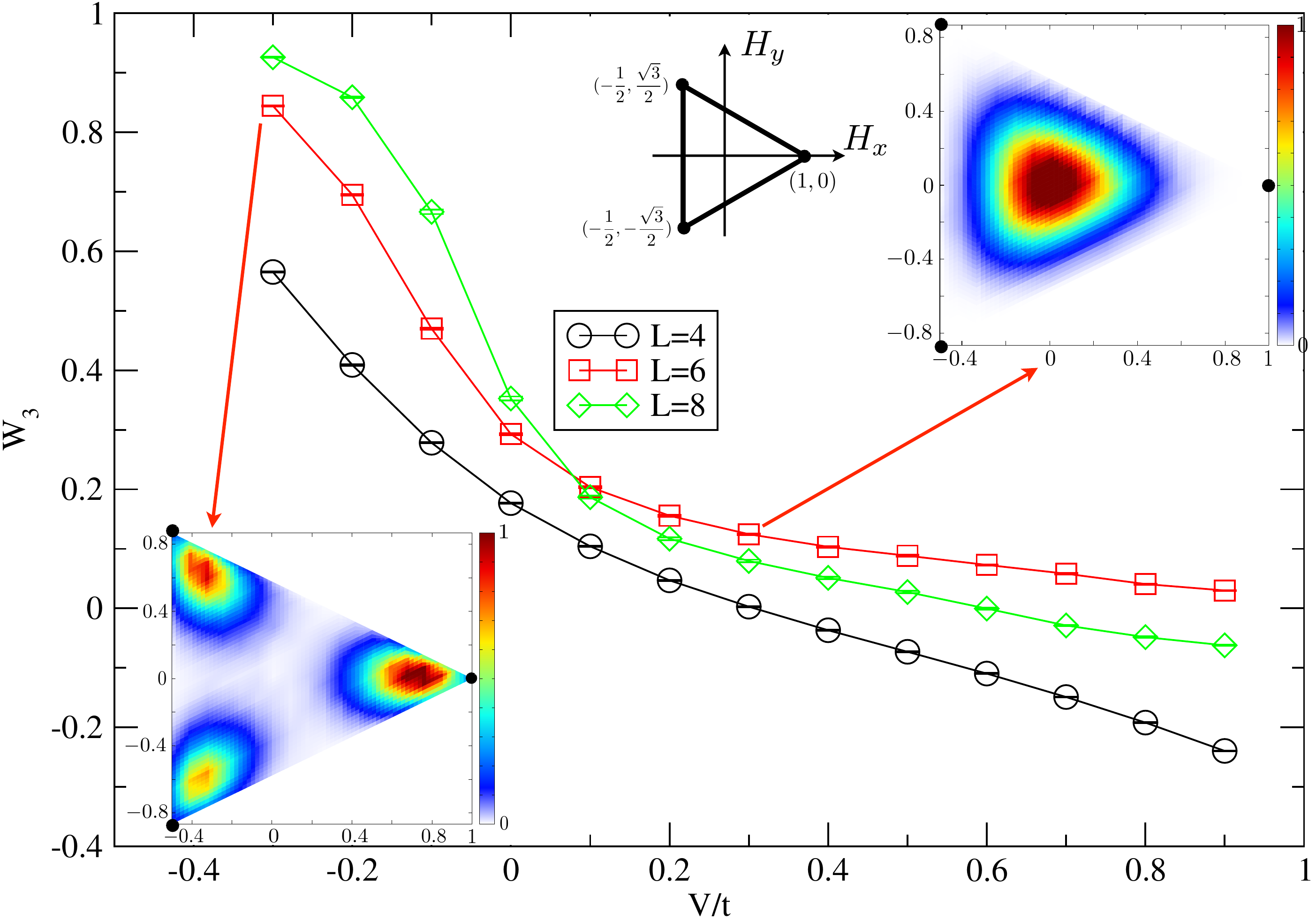}
\end{center}
\caption{(Color online) Expectation value $W_3$ of the
  quantifier of $Z_3$ anisotropy of histograms of loop occupations, as
  a function of $V/t$ for three even lattices sizes $L=4,6,8$. Insets:
  Color-coded normalized histograms of loop occupations numbers (using a color interpolation scheme), deep in the stripe phase (left inset, $L=6$, $V/t=-0.3$) and closer to the
  critical point (right inset, $L=6$, $V/t=0.3$). Vertical axis of histograms is $H_y$ (ranging from $-\sqrt{3}/2$ to
  $\sqrt{3}/2$), horizontal axis is $H_x$ (from $-1/2$ to $1$). The histograms are enclosed in an equilateral triangle pictured close to the right inset.}
\label{fig:loop_cos}
\end{figure}

Note that, in analogy with the situation in the QDM, one may suspect that other types of symmetry-breaking order (such as plaquette order) could also exhibit a non-zero value of $m_{\text{stripe}}$. To clarify this, it is useful to consider the histograms of the occupations of loop segments appearing in the Monte Carlo simulations, as was done in similar simulations for valence bond crystals~\cite{Sandvik07,Lou09,Albuquerque11,Sandvik12,Pujari13}. To have a two-dimensional representation of the histograms for the loop segments occupying one of the three lattice directions, we introduce $H_x=\frac{1}{2N} ( 2 N_0-N_1-N_2)$ and $H_y=\frac{\sqrt{3}}{2N}(-N_1+N_2)$ (with $N_\alpha$ being the total number of bonds in the direction $\alpha$). 
The three line configurations shown in the left part of Fig.~\ref{fig:loop_phase_diag} (having respectively $N_0,N_1$ 
or $N_2$ equal to $N$) occupy, in the $(H_x,H_y)$ representation, the three vertices $(1,0),(-1/2, \pm \sqrt{3}/2)$ of an equilateral triangle in which the histogram is enclosed, as represented in Fig.~\ref{fig:loop_cos}. The left inset of Fig.~\ref{fig:loop_cos} presents the histogram obtained for a moderate system size $L=6$ and $V/t=-0.3$, which is already located deep inside the stripe phase: we clearly observe a $\mathbb{Z}_3$-symmetric feature with three peaks (corresponding to high occupations) close to the corners of the triangle corresponding to the three line configurations. The distance of these peaks from the origin $(0,0)$ gives a measure of the order parameter $m_{\text{stripe}}$. 

On the other hand, the {\it angles} $\theta=\arctan(H_y/H_x)$ associated to the three peaks in this two-dimensional representation also carry important information on the precise nature of the phase. For instance, the angles associated with  the three stripe patterns are $0$, $2 \pi/3$ and $4\pi /3$. Other candidate phases (such as the equivalent of plaquette phases) would show up in the histograms as sharp peaks at different angles. We find that, as $V/t$ gets closer to the critical point $(V/t)_{\text{c}}$, the typical extents (radius) of the histograms get smaller (as expected from the decrease of the order parameter), and that at the same time the shape of the histograms change from a $\mathbb{Z}_3$-symmetric one to a circular-symmetric one as is exemplified in the right inset of Fig.~\ref{fig:loop_cos}. This is reminiscent of the $U(1)$ valence bond histograms observed in the vicinity of deconfined quantum critical points~\cite{Sandvik07,Lou09,Lou09b,Sandvik12} or even at moderate proximity of the RK point of the QDM on the square lattice~\cite{Banerjee2014,Schwandt2015}. 

To quantify this effect, we calculated a measure of the $\mathbb{Z}_3$-symmetry of the histograms $W_3=\langle \cos(3\theta) \rangle$ using formulations similar to those in Refs.~\cite{Lou07,Lou09,Lou09b,Sandvik12,Pujari13}. For a pure stripe phase, we have $W_3=1$, while for a circular histogram $W_3=0$. The variation of $W_3$ as a function $V/t$ with different system sizes
displayed in the main panel of Fig.~\ref{fig:loop_cos} confirms the behavior observed visually in the histograms: for the region $V/t \lesssim 0.1$, $W_3$ tends to approach unity with increasing system sizes, while in the region $ 0.1 \lesssim V/t \leq 1$, $W_3$ appears to vanish within the system sizes available to us. While therefore we cannot exclude an intermediate different crystalline phase, it appears unlikely as there is no direct evidence either in the energy or in the structure factor of a phase transition (see a similar situation for the square lattice QDM~\cite{Banerjee2014,Schwandt2015}).
Thus, we expect that on larger systems $W_3$ would also tend to approach unity  for $V/t \leq (V/t)_{\text{c}}$. Note that the computations of $W_3$ are difficult as we need long Monte Carlo runs to be fully ergodic ({\em i.e.,} in order not to get locked in one type of line configurations, in particular in the stripe phase), and also to ensure that statistical fluctuations are not too important when the order parameter is small. We may speculate that these circular histograms
are indications of the $\mathbb{Z}_3$ anisotropy being a dangerously irrelevant variable at the critical point $(V/t)_{\text{c}}$. Although the precise characterization of the transition between the $\mathbb{Z}_2$ spin liquid and the stripe phase is an interesting issue, it is beyond the scope of the present work. At the point $V/t=0$ of our main interest, our QMC data (see Figs.~\ref{fig:loop_sf} and~\ref{fig:loop_cos}) clearly indicate that the ground state of the effective Hamiltonian \eqref{eq:ring_hamiltonian} is located in the stripe phase. The resulting phase diagram obtained from these QMC results as well as the general considerations in Sec.~\ref{sec:loop_general} is given in Fig~\ref{fig:loop_phase_diag}.

\section{Numerical simulations of the microscopic model}
\label{sec:spin_model}
In this section, we present the results of the QMC simulations for the original spin-1/2 Hamiltonian
(\ref{eq:BFG_model}) at the $m=1/6$ plateau. We used the Stochastic Series Expansion (SSE) algorithm~\cite{Sandvik1991,Syljuasen2002} combined with a plaquette generalization~\cite{Louis2004} that is necessary to circumvent the freezing encountered in the bond formulation of the SSE when $J_{z}$ is large~\cite{Melko2006,Melko2007}. In addition to this difficulty,
capturing ground state physics requires to reach a very low temperature-range characterized by the small energy scale $T \sim J_{\mathrm{ring}} \sim J_{xy}^2/J_z$ of the effective model. Nevertheless, we have been able to simulate $L\times L$ systems, with $N=3L^2$ sites, up to $L\simeq 12$, or slightly beyond $L\simeq 18$ for some observables. For numerical ease, we restrict ourselves to the truncated model where only the nearest-neighbor 
interactions (i.e., the black bonds in Fig.~\ref{fig:kagome_lattice}) 
are retained in $H_{xy}$ (see the discussions in Sec.~\ref{sec:mapping-to-QDM}). 

\begin{figure}
\begin{center}
\includegraphics[width=\columnwidth,clip]{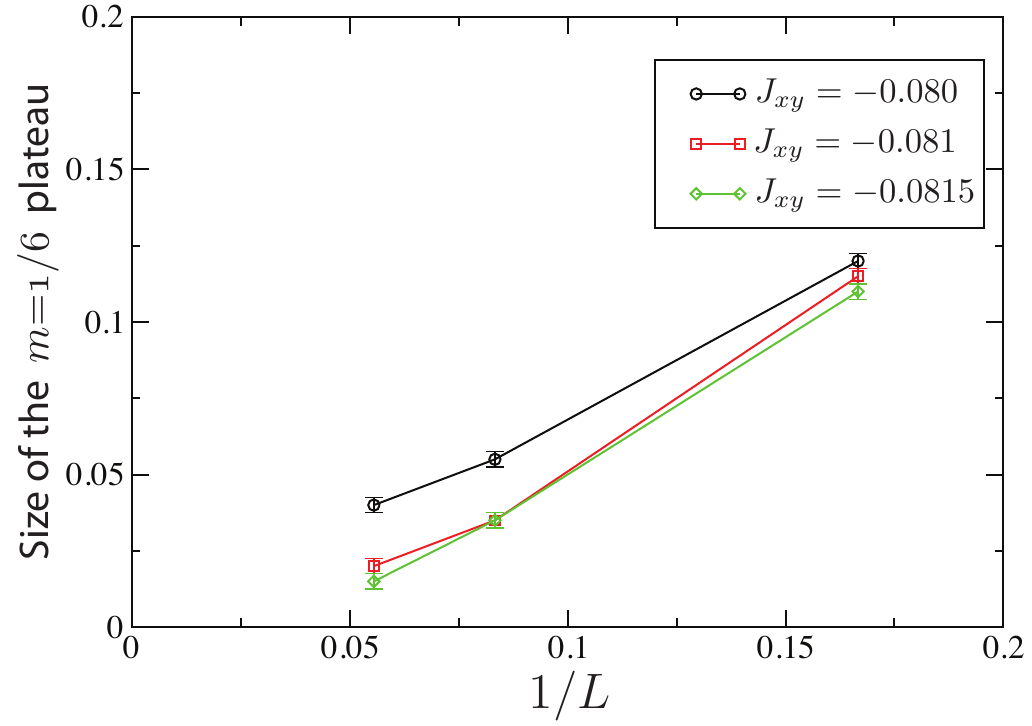}
\end{center}
\caption{(Color online) Finite-size scaling of the $m=1/6$ plateau width. The plateau is found to disappear around $J_{xy} \simeq - 0.081$.}
\label{fig:plateau_size}
\end{figure}

In the following, we set $J_z = 1$ as the energy scale, and use ferromagnetic values $J_{xy} <0$ (for the QMC simulations to have no sign problem) and $h=1.03$ for the magnetic field (corresponding to the middle of the $m=1/6$ plateau; see Fig.~\ref{fig:MofH}). In order to locate the transition point out of the plateau phase as a function of $J_{xy}$, we performed a finite-size analysis of its width. Resulting data are shown in Fig.~\ref{fig:plateau_size}, and the transition point is estimated to be around $J_{xy} \simeq - 0.081$. For values $J_{xy}< -0.081$, the system is thus in a planar phase (superfluid in the bosonic language), and in the following we focus on the plateau region $-0.081 < J_{xy} <0$. In Sec~\ref{sec:plateau_nature}, we examine the nature of the plateau phase, and discuss what could be the scenarios for the transition to the planar phase. Sec~\ref{sec:plateau_renyi} is devoted to an alternative characterization of the phase using the R\'{e}nyi entanglement entropy.

\subsection{Nature of the $m=1/6$ plateau}
\label{sec:plateau_nature}

For magnetization plateaus, as has been mentioned in the introduction, it is useful to consider the quantity $q(S-m)$, where $q$ is the number of sites per unit cell ($q=3$ for the Kagom\'e lattice). Indeed, it has been shown analytically~\cite{Oshikawa2000,Hastings2004} that when this quantity is fractional, the ground state on this plateau must be degenerate, {\em i.e.}, either in a crystalline state with magnetic superstructures or in a topological state, while a non-degenerate featureless state with a gap is forbidden. If, on the other hand, it takes an integer value, which is the case here 
at the $m=1/6$ plateau (see Table~\ref{tab:effective-model-summary}), all these possibilities are allowed.  

\begin{figure*}
\begin{center}
\includegraphics[width=0.9\textwidth,clip]{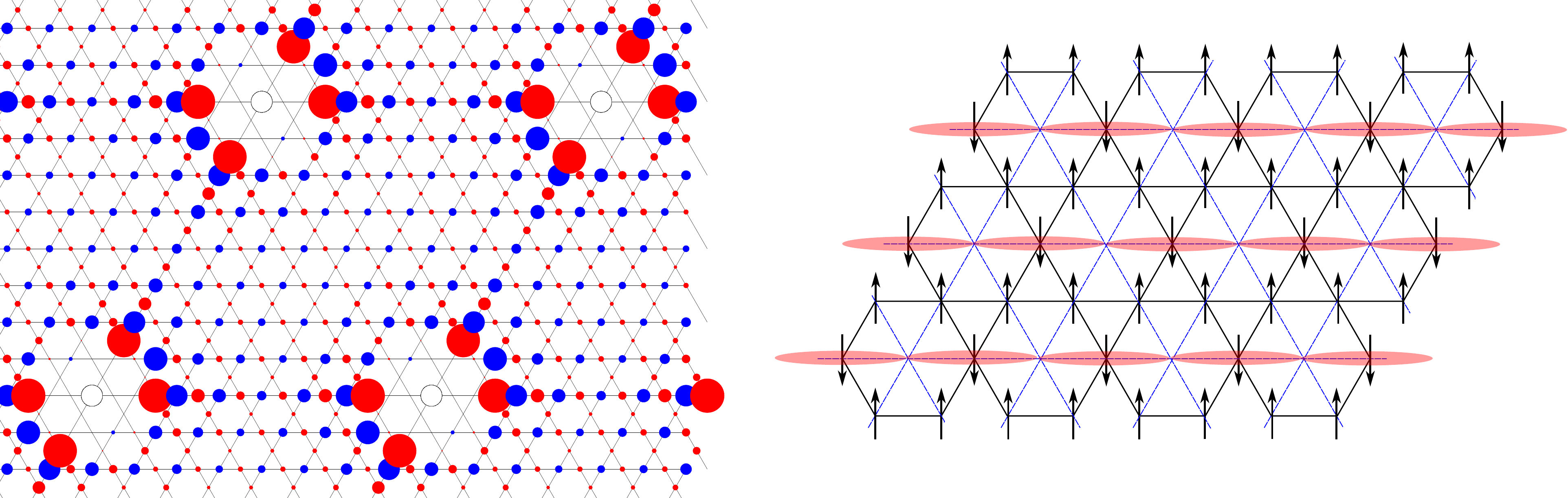}
\end{center}
\caption{(Color online) Left: Connected correlation function$\langle S^z_i S^z_j\rangle_{\text{c}}= \langle S^z_i S^z_j \rangle - \langle S^z_i \rangle \langle S^z_j \rangle $ on the $m=1/6$ plateau (for $J_{xy}=-0.078$, $T=J_{\mathrm{ring}}/8$ on a $L=8$ cluster with periodic boundary conditions). Reference site is indicated as a black circle. Blue (respectively red) denote positive (resp. negative) values, and the radius is proportional to the absolute value of the correlation. Data close to the reference site have not been plotted for readability. Right: Cartoon representation of the spin and dimer configurations in one state of the stripe phase, with the Kagom\'{e} lattice in black and the underlying triangular lattice in blue.}
\label{fig:QMCcorr}
\end{figure*}

Guided by the results of the effective QLM in Sec~\ref{sec:loop_model}, we expect the plateau phase to be in a ordered phase where the down spins form stripes. This is indeed evident when plotting the connected $S^z$-$S^z$ spin correlation function in real space, as shown in the left panel of Fig.~\ref{fig:QMCcorr}, where one readily sees the presence of stripes. The corresponding phase breaks the six-fold rotation symmetry, and is three-fold degenerate. Cartoon pictures of the spin and dimer configurations in this phase are drawn on the right panel of Fig.~\ref{fig:QMCcorr}.

\begin{figure}
\begin{center}
\includegraphics[width=\columnwidth,clip]{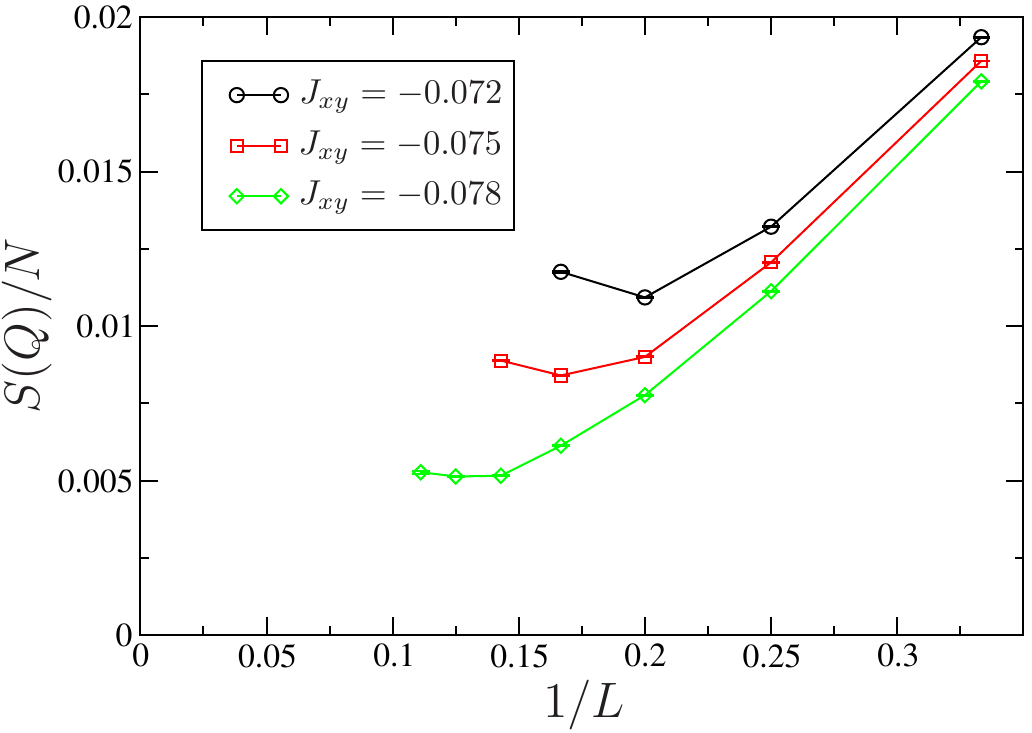}
\end{center}
\caption{(Color online) Finite-size scaling of the (scaled) structure factor $S({\mathbf Q})/N$ inside the magnetization plateau. Despite the relatively small available sizes, we can reach the regime of saturation, confirming the striped solid anticipated from the QLM.}
\label{fig:scalingSQ}
\end{figure}

In order to confirm the long-range nature of this order, we compute the Fourier transform of the above spin-spin 
correlations, {\it i.e.}, the diagonal spin structure factor:
\begin{equation}
S({\mathbf q}) = \frac{1}{N}\sum_{j,k} ( \langle S^z_j S^z_k \rangle - \langle S^z_j \rangle \langle S^z_k \rangle) \mathrm{e}^{i{\mathbf q} \cdot ({\mathbf r}_j-{\mathbf r}_k)},
\label{eq:structure_factor_scaling}
\end{equation}
where ${\mathbf r}_j$ is the position of spin $j$, $N=3L^2$, ${\mathbf q}=(q_x,q_y)$ and the average local magnetization is $\langle S^z_j \rangle = m = 1/6$. For an ordered phase at wave-vector ${\mathbf Q}$, $S({\mathbf Q})$ diverges as $N$, and for the three-fold degenerate solid, the three Bragg peaks are located at wave-vectors ${\mathbf Q}_1=(0,4\pi/\sqrt{3})$, ${\mathbf Q}_2=(2\pi,2\pi/\sqrt{3})$ and ${\mathbf Q}_3=(2\pi,-2\pi/\sqrt{3})$. Note that this solid can be accommodated on any cluster used for the simulations. Finite-size scaling of $S({\mathbf Q})/N$ as a function of $1/L$ for several values of $J_{xy}$ in the plateau phase clearly indicates the presence of the stripe phase (see Fig.~\ref{fig:scalingSQ}). In agreement with the QLM study at $V/t=0$, the order develops already for small system sizes.

Two comments are in order here. First, $S({\mathbf Q})$ used in Fig.~\ref{fig:scalingSQ}
is the average of the values at the three wave-vectors ${\mathbf Q}_{\alpha}$, which is valid when the simulation is perfectly ergodic. However, as mentioned earlier, the large anisotropy $J_z/J_{xy} \sim 13$ in the parameter range of interest can create difficulties in the QMC simulations~\cite{Melko2007}.  
By monitoring the three values $S({\mathbf Q}_{\alpha})$ separately, we have checked that the simulations are ergodic for the sizes $L\leq 9$ considered here. However, for smaller values
of $|J_{xy}|/J_z$ or larger $L$, we indeed noticed a quick loss of
ergodicity, even with the plaquette algorithm, translating into very
different values at the three wave-vectors. The same behavior was observed in Ref.~\onlinecite{Melko2006} which used a similar QMC algorithm. 
Second, we sometimes found a non-monotonic behavior of the structure factor $S({\mathbf Q})/N$ for larger sizes (see Fig.~\ref{fig:scalingSQ}). Again,
a similar behavior was reported in Ref.~\onlinecite{Melko2006} where the increase of the order parameter was interpreted as the threshold where stripes start developing in the system. In
practice, this complicates a proper extrapolation of the order parameter 
to its thermodynamic limit value. We can however compare the order of magnitude
obtained with the value for a perfectly equal superposition of the three
ordered states, for which a straightforward calculation yields
$S({\mathbf Q})/N=8/81\simeq 0.0987$.

We conclude this section by a discussion on the possible scenarios for the quantum phase transition between the plateau (stripe) crystal and planar phases. Since the stripe crystal breaks discrete lattice symmetries, one expects, on general grounds, that the transition will be of first order type, or that both phases coexist (in a small region of the phase diagram). This latter scenario implies the existence of an intervening supersolid phase, whose presence was found in several similar models on other frustrated lattices~\cite{Wessel2005,Heidarian2005,Melko-SS-2005}. Due to the limited systems sizes available to us, we have not been able to favor one of these two scenarios. We observed (data not shown) an apparent crossing in the spin stiffness, with critical exponents compatible with a continuous phase transition, and no sign of a double peak structure in the kinetic energy histograms. However, it has been shown in similar situations that the correct nature of the phase transition may be difficult to capture. In the ($J_{xy}$,$h$) plane, varying the XY interaction $J_{xy}$ at the constant field $h=1.03$ corresponds to a transition point located close to the tip of plateau phase ({\it i.e.} the tip of the insulator lobe in the bosonic picture), for which previous studies have revealed that such a crystal-planar phase transition may appear to be continuous while it is in fact \emph{weakly first-order}~\cite{Isakov2006a,Damle2006,Arnab2007}. We believe that such a picture should also apply here, although proving this definitively would imply simulations on very large samples, which is currently out of reach for the present model.

\subsection{R\'enyi entropy in the crystalline phase}
\label{sec:plateau_renyi}
We have shown, using a conventional approach ({\it i.e.} by guessing a broken symmetry and then measuring the corresponding structure factor), that the $m=1/6$ plateau corresponds to a stripe crystal 
shown in Fig.~\ref{fig:QMCcorr}. It is interesting to check whether other means could detect the nature of the ground state without a priori knowledge on the broken symmetry. An elegant approach to obtain the dominant order parameter in an unbiased way is provided by the correlation density matrix~\cite{Cheong2009}, but it is not easily accessible within QMC simulations.  Instead, we will focus here on using the entanglement entropy as a means to access the nature of the ground state.

Indeed, in the recent years, a large number of studies have shown how the scaling behavior of a block entanglement entropy can give access to some ground state properties, such as the central charge in one-dimension~\cite{Calabrese2004} or the topological order underlying a given ground state wave function~\cite{Levin2006,Kitaev2006}. Generically, any entanglement entropy (whether R\'enyi or von Neumann) will scale with the ``area'' of the boundary between the block and the rest of the system (the so-called ``area-law''; see Ref.~\onlinecite{Eisert2010} for a review), possibly with interesting (universal) sub-leading terms:
\begin{equation}\label{eq:Sq}
S_{q}(\text{A}) = a_q\ell_{\text{A}} + d_q
\end{equation} 
where $q$ is the R\'enyi index, $\ell_{\text{A}}$ the length of the boundary of block A and $d_q$ a constant for instance. Topological phases can be detected using the constructions of Levin-Wen~\cite{Levin2006} or Kitaev-Preskill~\cite{Kitaev2006} (see also Ref.~\onlinecite{Furukawa2007}), that enables one to extract a negative constant term $d_q = -\gamma < 0$, independent of $q$~\cite{Flamina2009}, where $\gamma$ is the topological entanglement entropy.  This was used for instance to conclude that the ground state of the BFG model at $m=0$ is indeed a  $\mathbb{Z}_2$ spin liquid with $\gamma=\log 2$
~\cite{Isakov2011}.

\begin{figure}
\begin{center}
\includegraphics[width=\columnwidth,clip]{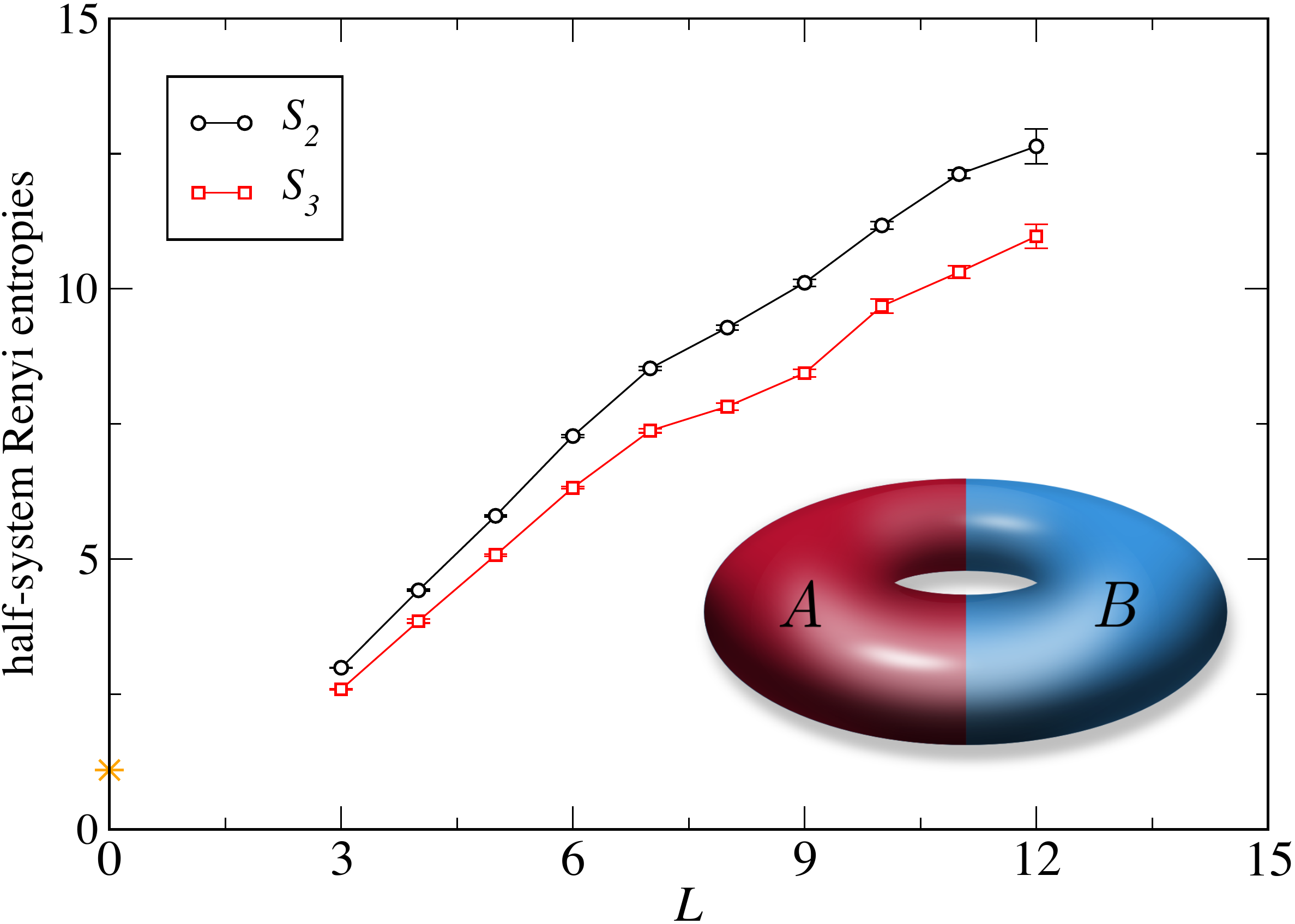}
\end{center}
\caption{(Color online) R\'enyi block entanglement entropies $S_q(\text{A})$ ($q=2,3$) on the $m=1/6$ plateau, as a function of $L$, on various systems with $N=3L^2$ sites. Simulations parameters are $J_{xy}=-0.078$, $T=J_{\mathrm{ring}}/4$ and $h=1.03$. In each case, the A block corresponds to a half-system cylinder containing $(L/2)\times L$ unit cells (see inset). 
The positive correction $\log (3)$ is shown by a star.}
\label{fig:Renyi_vs_L}
\end{figure}

The constructions of Refs.~\cite{Levin2006,Kitaev2006} use a subtraction of terms which all scale like the area of the subsystems considered, which can be troublesome in QMC simulations which inevitably exhibit finite error bars. Computing the scaling of $S_q(\text{A})$ for a single block A (without using subtraction schemes), is in general computationally simpler. This is what we performed for the case of our discrete symmetry breaking ($C_{6\text{v}}\mapsto C_{2\text{v}}$) on the $m=1/6$ plateau, taking for A the geometry of a cylinder block embedded in a torus sample (see the inset of Fig.~\ref{fig:Renyi_vs_L}).

We expect that for such a symmetry-broken state, the sub-leading term will be a \emph{positive} constant $d_q = \log g$, where $g$ is the number of degenerate ground states (note the sign of $d_{q}$; see Appendix~\ref{sec:appendixA}). We first remark that, for the simple type of order on the $m=1/6$ plateau, the precise block geometry is not very important, which allows us to consider the convenient cylinder block geometry. Second, it is quite important for this result to hold that all $g$ degenerate states (quasi degenerate on a finite system but with an exponentially small splitting) are observed in the QMC simulations. Hence, ergodicity must be ensured to extract the degeneracy, which we observe is the case if we are not too deep inside the stripe phase and system sizes are not too large. Note that such a {\em positive} correction gives a vanishing contribution to the previously mentioned geometric constructions~\cite{Levin2006,Kitaev2006}, as they are precisely built to solely capture long-range entanglement.
Our numerical data are presented in Fig.~\ref{fig:Renyi_vs_L}. We have
computed the R\'enyi entropies $S_2$ and $S_3$ on the $m=1/6$
plateau at low temperatures using the QMC extended ensemble method proposed in Ref.~\onlinecite{Humeniuk2012} (with a small modification where the size of A is dynamically varied during the simulation). 

An important first observation is that if one considers only the smallest (four or five) system sizes, a naive fit leads to a negative intercept and thus to an apparent
topological phase. One could perhaps interpret this unexpected finite-size effect in the following way: close enough to the transition to the superfluid phase, the superfluid correlation length can be of the order of, or larger than, the system size for small enough samples. Effectively, the system acquires a superfluid component, which we expect to contribute to $S_q$ as $a'_q \ell_A + b \log  ( \ell_A) + d'_q$.
This form is derived in Ref.~\onlinecite{Metlitski2011} for a system that spontaneously breaks a continuous symmetry, in which case the logarithm prefactor $b$ is universal (positive and related to the number of Goldstone modes), and $d'_q$ a non-universal constant. Assuming such a form is also valid in a system which has effectively both superfluid and crystalline correlations ({\it i.e.}, supersolid), it could result in fits that display a non-universal, possibly negative (for small enough systems, since otherwise $b \log ( \ell_A) \gg |d'_q|$ for large enough $L$), constant intercept. Note that it would be interesting to check this behaviour using simpler bosonic models that exhibit supersolid behaviour.

However, when the system is large enough (beyond $L=7$ in our simulations), we observe a qualitative change of behavior in $S_q$ as a function of $L$. Note that this length scale coincides with the one where the stripe structure factor starts saturating, as observed in Fig.~\ref{fig:scalingSQ}. For these larger systems, entanglement entropies are compatible with a positive intercept $d_q \simeq \log(3)$, although the accuracy is not excellent. For instance, fitting our $S_2$ data using (\ref{eq:Sq}) and sizes with $7 \leq L \leq 10$, we get $d_q\simeq 0.9$, which deviates quite significantly from $\log(3)$. Indeed, as already advocated, a precise determination of $d_q$ is rather difficult due to the need of excellent ergodicity and hence large autocorrelation times for this quantity. Typically, we need more than $10^8$ measurements to get reliable data.

\section{Conclusion}
\label{sec:conclusion}
In search for magnetization plateaus of spin-liquid nature, we have considered the effect of a magnetic field on an anisotropic spin-1/2 model on the Kagom\'e lattice that was introduced in Ref.~\onlinecite{Balents2002}. In the easy-axis limit, this model is designed to exhibit several magnetization plateaus, at magnetization per site $m=0, 1/6, 1/3$ (plus saturation $1/2$). The properties of these plateaus have been investigated first by mapping the original spin model to simpler constrained models on the triangular lattice. Focusing on the $m=1/6$ plateau, we have investigated in detail the effective quantum loop model and mapped out its phase diagram using a reptation quantum Monte-Carlo algorithm. From this, we predict a 3-fold degenerate stripe phase, that breaks rotation symmetry, to appear on this magnetization plateau.  
We have also observed that the $m=1/6$ plateau with the stripe order turns into 
a topological $\mathbb{Z}_{2}$ spin-liquid plateau when an additional repulsion among 
the dimer segments is added. 

Then, we have performed large-scale numerical simulations using the SSE
quantum Monte-Carlo algorithm on the original microscopic model. This
has confirmed the existence of a stripe phase found in the effective loop 
model, using both standard structure-factor measurements and its
signature in entanglement entropies. In particular, we have observed
that, for length scales shorter than $L\simeq 6$, the scaling of the block entanglement
entropy may be misleadingly interpreted as giving a \emph{negative} intercept $d_q$, while it becomes clearly
\emph{positive} for larger systems. This should be used as a caveat in
other situations.

When the XY-interaction $|J_{xy}|$ increases, there is a quantum phase transition to a planar phase 
(superfluid of the equivalent hardcore bosons). While it appears to be continuous in our simulations, we believe that it should appear weakly first-order on larger lattices, as observed for instance in similar physical situations~\cite{Isakov2006a,Damle2006,Arnab2007}.

We have not investigated in details the $m=1/3$ plateau, where the effective model is given by the standard quantum dimer model (\emph{i.e.} a single dimer per site) on the triangular lattice, that has been widely studied  in the past~\cite{Moessner2001,Ralko2005,Syljuasen2005,Ralko2006,Vernay2006,Ralko2007,Misguich2008a,Herdman-W-11}. In fact, the expected crystal in this phase would have a $\sqrt{12}\times \sqrt{12}$ unit cell and an extremely small order parameter~\cite{Ralko2006}. Given that numerical simulations of the microscopic spin model are more involved (larger systems, many different energy scales etc.), it remains challenging to detect such a weak order in the direct simulations for the original BFG spin model. 

Before concluding, let us mention possible extensions of the model (\ref{eq:BFG_model}).  
In Sec.~\ref{sec:loop_model}, we have seen that there exists a gapped $\mathbb{Z}_{2}$ spin liquid phase 
around the RK point of the QLM (see Fig.~\ref{fig:loop_winding}). Therefore, it would be interesting to drive the system toward the spin-liquid phase.  
In fact, it is possible to consider additional interactions which would result in a non-zero interaction term $V$ in the effective constrained model (\ref{eq:QDM_triangular}). The four-spin interaction $Q$ [Eq.~\eqref{eqn:4-spin-int}] or the inter-hexagon third-neighbor interaction $J_{z}^{(3)}$ [Eq.~\eqref{eqn:inter-hex-3rdN}], that is not included in the model \eqref{eq:BFG_model}, should do the job.
This could allow us to reach the $\mathbb{Z}_2$ liquid state at both $m=1/3$ and $m=1/6$ magnetization plateaus 
when $Q, J_{z}^{(3)} \sim 2J_{xy}^{2}/J_{z}$, 
while the transition between the crystalline and topological phases could in principle be monitored by examining the R\'enyi entropy behavior. This extension of the BFG model would certainly be a very interesting playground to investigate such topological phases, their detections and their transitions to conventional symmetry-breaking states. 
Also our construction is not restricted to spin-1/2 systems (or to the equivalent hardcore-boson systems) 
and applies to higher-spin systems as well.  
In fact, we can obtain the same effective models (QDM and QLM) for the spin-1 version of 
the model \eqref{eq:BFG_model} with an additional single-ion anisotropy $D\sum_{i}(S_{i}^{z})^{2}$.    

As a final remark, let us briefly comment on possible experimental realization of the BFG model.  
Because of the artificial interactions required to build the ice manifold 
of the BFG model, it is difficult to make direct connections between the model we considered and  
the existing Kagome compounds \cite{Mendels-W-review-11}.  
However, it remains an interesting prospect to investigate whether such toy models could be realized experimentally 
in tunable artificial systems. Indeed, it was suggested very recently that the BFG interactions 
and hexagonal plaquettes could be reproduced using a 2D cold ion crystal, with an example 
given for one and two hexagons~\cite{Nath2015}.

{\em Note added:} While finishing this manuscript, we became aware of the recent preprint by Roychowdhury et al.~\cite{Roychowdhury2015} who recently derived the same loop model for the equivalent hardcore-boson Hamiltonian, 
and further studied the phase diagram of the loop model with a potential term. While we agree on the phase 
structure of the loop model, the position of the critical point separating the stripe from 
the $\mathbb{Z}_2$ liquid phase is noticeably different 
[$(V/t)_{\text{c}} \sim -0.3$ versus $(V/t)_{\text{c}} \sim 0.7$ ].

\begin{acknowledgments}
We acknowledge useful discussions with G. Misguich, R. Melko, and D. Schwandt.  
We also thank F. Pollmann for the helpful communication on their results. 
This work was performed using HPC resources from GENCI (grants x2014050225 and x2015050225) and CALMIP (grants 2014-P0677 and 2015-P0677). Our QMC SSE simulations are based on the code from the ALPS libraries~\cite{ALPS2}. One of the authors (K.T.) was supported in part by JSPS KAKENHI Grant No. 24540402 and No. 15K05211
\end{acknowledgments}

\appendix
\section{R\'enyi entropy in a discrete symmetry breaking phase} \label{sec:appendixA}
Let us consider $g$ states $\psi_i$, each corresponding to one of the degenerate states in the thermodynamic limit. This would be for instance $|\uparrow \downarrow \uparrow \downarrow \ldots\rangle$ and $|\downarrow \uparrow \downarrow \uparrow \ldots\rangle$  in an Ising phase, or the three states with down spins occupying one of the three Kagom\'e sublattices (and up spins on the other sites, see Fig.~\ref{fig:QMCcorr}), in the stripe crystal phase discussed in the main text. These $g$ states are always orthogonal in the thermodynamic limit. 
Since they are simple product states, one can readily show, for any block $A$ of the system which complement scales with system size, that the reduced density matrix corresponding to a cat superposition of these states, $|\psi_{\mathrm{cat}}\rangle = (1/\sqrt{g}) \sum_i |\psi_i\rangle$, is given by:
\begin{equation}\label{eq:cat}
\hat{\rho}^A_{\mathrm{cat}} = \frac{1}{g} \sum_i \rho^A_i. 
\end{equation}
Note that this equality only holds for the reduced density matrix, and not the whole density matrix. 

On any finite system, since these $g$ states are exponentially close in energy, and a QMC simulation would sample each $|\psi_i\rangle$ with equal probabilities. Hence, one would compute a reduced density matrix corresponding to a mixed state:
\begin{equation}
\hat{\rho}_{\mathrm{QMC}} = \frac{1}{g} \sum_i \rho_i 
\quad \Longrightarrow \quad 
\hat{\rho}^A_{\mathrm{QMC}} = \frac{1}{g} \sum_i \rho^A_i 
\end{equation}
which is precisely the same object as Eq.~(\ref{eq:cat}). So our QMC simulation is able to access the reduced density matrix of the cat state, provided it is ergodic. 

Now, each $\rho^A_i$ is excessively simple for each product state, as its spectrum contains only one non-zero eigenvalue $\lambda=1$ since they are not entangled. Moreover, they can be diagonalized simultaneously (they are diagonal in the $S^z$ basis in our example), which leads to an entanglement spectrum with non-zero eigenvalues $1/g$ with degeneracy $g$. From it, the entanglement entropy (von Neumann or Renyi) is simply $\log g$. 

This results looks quite intuitive indeed and was conjectured in Ref.~\cite{Furukawa2007}. It was already noticed in Ref.~\cite{Jiang2012} for a cat state made of $g$ degenerate Ising ground states. 
We emphasize that the same result holds for the mixed state obtained in our QMC simulation. 
For more realistic wave-functions, the fluctuations around the ideal product state gives and area law, but the constant term will be unchanged:
\begin{equation}
S_{q}(A) = a_q\ell_A + \log g + \ldots
\label{eq:renyi_scaling_crystal}
\end{equation}

As a last remark, we would like to emphasize that in other more complex cases (e.g. for states which are not one-site product states, such as valence-bond columnar states), the discussion is more involved as the constant term can now depend explicitly on the form of the cut as well as on the R\'enyi index $q$.



%

\end{document}